\documentclass[journal]{IEEEtran}
\usepackage{mathrsfs}

\usepackage{verbatim}
\usepackage{makecell}
\usepackage{bbm}
\usepackage{amssymb}
\usepackage[caption=false]{subfig}
\usepackage[mathcal]{euscript}
\usepackage{float}
\usepackage{psfrag,calc,url,bm}
\usepackage{cite}
\usepackage{graphicx}
\usepackage{psfrag}
\usepackage{array} 
\usepackage{url}
\usepackage{stfloats}
\usepackage{amsmath}
\usepackage{float}
\usepackage{algorithmic}
\usepackage[ruled,linesnumbered,vlined]{algorithm2e}
\usepackage{color}
\usepackage{boxedminipage}
\usepackage{amsthm}
\usepackage{multirow}
\usepackage{setspace}
\usepackage{soul}
\usepackage{epstopdf}
\allowdisplaybreaks[3]
\usepackage{bbding}
\usepackage{threeparttable}
\usepackage{booktabs} 
\usepackage{tabularx}

\begin{document}


\title{Agentic Graph Neural Networks for Wireless Communications and Networking Towards Edge General Intelligence: A Survey}



\author{Yang Lu,~\IEEEmembership{Member,~IEEE}, Shengli Zhang, Chang Liu, Ruichen Zhang,~\IEEEmembership{Member,~IEEE}, Bo Ai,~\IEEEmembership{Fellow,~IEEE},\\
Dusit Niyato,~\IEEEmembership{Fellow,~IEEE},~Wei Ni,~\IEEEmembership{Fellow,~IEEE},~Xianbin Wang,~\IEEEmembership{Fellow, IEEE},~Abbas Jamalipour,~\IEEEmembership{Fellow,~IEEE}
\thanks{Yang Lu, Shengli Zhang, and Chang Liu are with the State Key Laboratory of Advanced Rail Autonomous Operation, and  also with the School of Computer Science and Technology, Beijing Jiaotong University, Beijing 100044, China (e-mail: \{yanglu, 24140039, 21281042\}@bjtu.edu.cn).}
\thanks{Ruichen Zhang and Dusit Niyato are with the College of Computing and Data Science, Nanyang Technological University, Singapore 639798 (e-mail: \{ruichen.zhang, dniyato\}@ntu.edu.sg).}
\thanks{Bo Ai is with the School of Electronics and Information Engineering, Beijing Jiaotong University, Beijing 100044, China (e-mail:  boai@bjtu.edu.cn).}
\thanks{Wei Ni is with Data61, CSIRO, Marsfield, NSW 2122, Australia, and the
School of Computing Science and Engineering, and the University of New South Wales, Kensington, NSW 2052, Australia (e-mail: wei.ni@ieee.org).}
\thanks{Xianbin Wang is with the Department of Electrical and Computer Engineering, Western University, London, Canada N6A 5B9 (emails:
xianbin.wang@uwo.ca).}
\thanks{Abbas Jamalipour is with the School of Electrical and Computer Engineering, University of Sydney, NSW 2006, Australia (e-mail: a.jamalipour@ieee.org).}
}

\maketitle

\begin{abstract}
The rapid advancement of communication technologies has driven the evolution of communication networks towards both high-dimensional resource utilization and multifunctional integration. This evolving complexity poses significant challenges in designing communication networks to satisfy the growing quality-of-service and time sensitivity of mobile applications in dynamic environments. Graph neural networks (GNNs) have emerged as fundamental deep learning (DL) models for complex communication networks. GNNs not only augment the extraction of features over network topologies but also enhance scalability and facilitate distributed computation. However, most existing GNNs follow a traditional passive learning framework, which may fail to meet the needs of increasingly diverse wireless systems. This survey proposes the employment of agentic artificial intelligence (AI) to organize and integrate GNNs, enabling scenario- and task-aware implementation towards edge general intelligence. To comprehend the full capability of GNNs, we holistically review recent applications of GNNs in wireless communications and networking. Specifically, we focus on the alignment between graph representations and network topologies, and between neural architectures and wireless tasks. We first provide an overview of GNNs based on prominent neural architectures, followed by the concept of agentic GNNs. Then, we summarize and compare GNN applications for conventional systems and emerging technologies, including physical, MAC, and network layer designs, integrated sensing and communication (ISAC), reconfigurable intelligent surface (RIS) and cell-free network architecture. We further propose a large language model (LLM) framework  as an intelligent question-answering agent, leveraging this survey as a local knowledge base to enable GNN-related responses tailored to wireless communication research.   Finally, we highlight the critical challenges, open issues, and future research directions for GNN-empowered wireless designs.
\end{abstract}


\section{Introduction}

\subsection{Background}

With the rapid proliferation of deep learning (DL) across diverse application domains, we have entered the era of artificial intelligence (AI) \cite{10601684}. Typically, data-driven DL can automatically extract complex patterns and make intelligent decisions\cite{9398576}. Integrating AI into mobile communication systems has become an irreversible trend \cite{8808168}, aiming to boost the system performance via network-level intelligence and spur novel mobile intelligent applications \cite{10417099}. Numerous base stations (BSs) or access points (APs) are equipped with powerful computational resources either locally or via mobile edge computing (MEC) nodes \cite{9060868}. Consequently, mobile communication systems are undergoing rapid evolution toward edge general intelligence (EGI) \cite{10876185}, which leverages AI to empower mobile applications with advanced understanding and reasoning capabilities. Moreover, agentic AI has emerged as a robust paradigm featuring integrated capabilities in perception, reasoning, and action. The transition from traditional AI to agentic AI facilitates the shift from explicit command responses to proactive and goal-directed behavior. For instance, it is capable of making multi-stage decisions by adhering to human-like decision loops, such as the observe–orient–decide–act (OODA) loop \cite{9999129}. This capability is particularly suitable for wireless systems, given the time-varying channel conditions\cite{11061807}, service requirements\cite{9928395}, and network topologies. Agentic AI can also actively collect new datasets related to unseen wireless scenarios to enable continuous learning, thereby improving the models' expressive power. Therefore, agentic AI is expected to play a crucial role in enhancing mobile communication systems to realize EGI\cite{9745424}. 

\begin{figure*}[t]
\begin{center}
{\includegraphics[ width=1\linewidth]{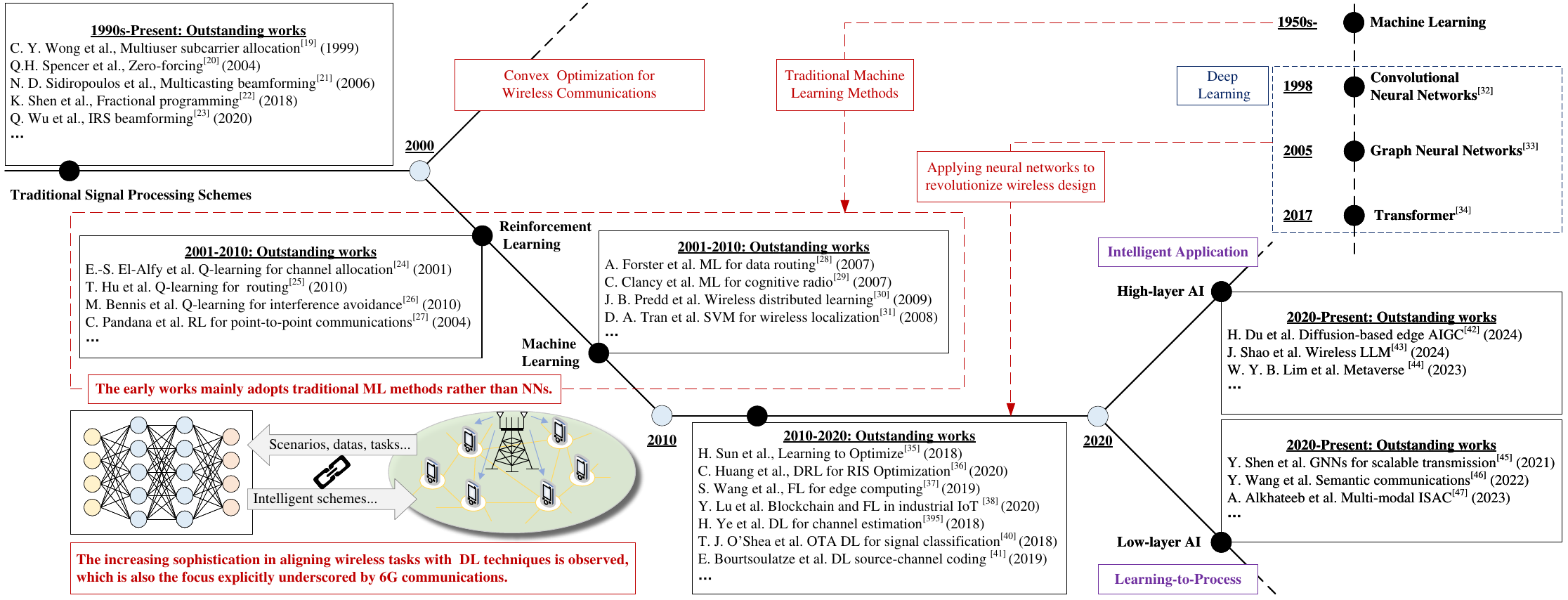}}
\caption{Timeline of the evolution from theoretical methods to DL approaches in wireless research, with two key observations: 1) Both mobile communication systems and AI undergo rapid development; 2) The alignment between wireless tasks and AI tools is increasingly strengthened.}
\label{Timeline}
\end{center}
\end{figure*}

Meanwhile, as the foundational infrastructure supporting the Internet of Everything (IoE), mobile communication systems possess growing service capabilities, benefited by recent hardware innovations and sophisticated waveform designs\cite{10422712}. Particularly, flexible-antenna technologies, such as reconfigurable intelligent surface (RIS)\cite{8741198}, also known as (a.k.a.) intelligent reflecting surface (IRS)\cite{8811733}, and fluid antenna system (FAS)\cite{9264694}, enable the creation of smart radio environments\cite{9140329}, and integrated sensing and communication (ISAC) \cite{9737357} facilitates the realization of multifunctional wireless services \cite{10562043}. Notably, future systems will integrate several novel techniques to enable diverse services. These techniques may be empowered by specialized DL models, and thus, agentic AI needs to coordinate these DL models to address complex tasks. Moreover, future communication systems confront unprecedented challenges, as novel hardware and network architectures complicate signal processing. Such a limitation not only impedes theoretical algorithms from attaining optimal performance, but also widens the performance gap between idealized outcomes and practical implementations. Therefore, optimizing wireless system performance remains a fundamental issue in both academia and industry. 

In light of the emerging trend toward empowering wireless systems with AI techniques, we present a brief timeline in Figure \ref{Timeline} to trace the evolution from theoretical methods to DL approaches in wireless research. The diverse communication scenarios and tasks have motivated numerous studies since the 1990s, particularly based on convex optimization \cite{timeline_boyd_1}, for example, multi-user subcarrier allocation\cite{timeline_wong_1}, zero-forcing (ZF)\cite{1261332}, multicasting beamforming\cite{1634819}, fractional programming \cite{8314727}, RIS transmission\cite{8930608}. The development of AI, especially machine learning (ML), has revolutionized many application fields since the 1950s. The application of ML to wireless networks can be traced to the early 21st century. Early studies explored reinforcement learning (RL) methods, such as Q-learning for channel allocation\cite{966366}, routing\cite{5408367} and interference avoidance\cite{5700414}, and RL for point-to-point communications\cite{1378063}, while others adopted supervised/unsupervised learning approaches, such as ML for data routing \cite{4496871} and cognitive radio\cite{4300983}, wireless distributed learning \cite{4802309}, and support vector machines (SVM) for wireless localization \cite{4384476}. These early works mainly employed traditional ML methods rather than neural networks (NNs). The rapid advancement of DL techniques including convolutional NNs (CNNs) \cite{726791}, graph NNs (GNNs) \cite{firstgnn}, and Transformer \cite{Vaswani_ANIP_2017}, has significantly propelled AI applications, spurring the development of DL-enabled wireless designs, such as learning to optimize\cite{learing-to-optimize}, deep RL (DRL) for RIS optimization \cite{9110869}, federated learning (FL) for edge computing \cite{8664630}, Blockchain for privacy in industrial Internet of Things (IoT)\cite{8843900}, and DL for channel estimation \cite{8052521}, signal classification\cite{8267032}, and source-channel coding\cite{8723589}. Since the 2020s, AI has been applied cross diverse network layers, enabling high-layer intelligence, including wireless AI-generated content (AIGC) \cite{10529221}, large language models (LLMs)\cite{10582827}, Metaverse\cite{9815180}, and low-layer intelligence including scalable resource allocation\cite{Shen_jsac_21}, semantic communications \cite{9832831} as well as  multi-modal sensing and communication\cite{10144504}.

\begin{figure*}[t]
\begin{center}
{\includegraphics[ width=1\linewidth]{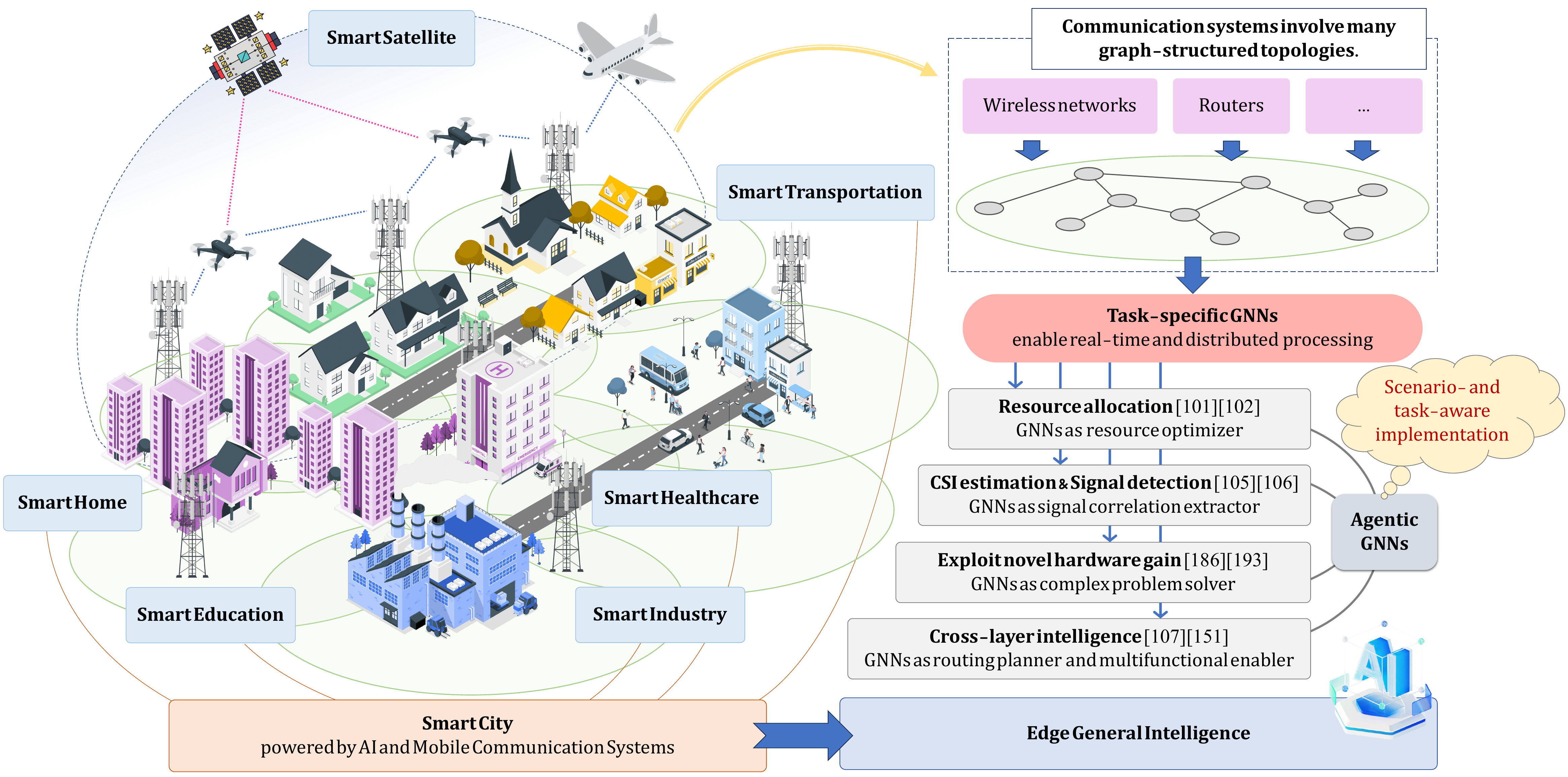}}
    \caption{An overview of agentic GNNs for wireless communications and networking towards EGI. It illustrates a smart city ecosystem powered by AI and mobile communication systems, encompassing interconnected scenarios (Smart Home, Healthcare, Transportation, etc.) and terrestrial and aerial elements (vehicles, UAVs, satellites, etc.). Communication systems involve numerous graph-structured topologies across layers, which can be enhanced by task-specific GNNs. By organizing GNNs in an agentic AI framework, agentic GNNs enable scenario- and task-aware implementation to support multiple functions, including resource allocation, CSI estimation and signal detection, exploitation of novel hardware gains, and cross-layer intelligence.}
\label{sys2}
\end{center}
\end{figure*}

This timeline highlights a discernible trend, i.e., the increasing sophistication in aligning wireless tasks with  DL techniques, which is also the focus explicitly underscored by the sixth-generation (6G) communications. The task-oriented nature of DL necessitates the customization of NNs for wireless communication systems\cite{ai_6g}. GNNs offer significant advantages over other models in tasks involving graph-structured data, such as cellular vehicle-to-everything (C-V2X) networks \cite{8967260}, device-to-device (D2D) communication networks \cite{9488278}, FL for edge computing \cite{10753492} and unmanned aerial vehicle (UAV) networks\cite{7470933}. GNNs have emerged as fundamental frameworks for addressing wireless-related tasks \cite{shengnn22}. By casting GNNs as a network topology-aware feature extraction module, they can be seamlessly integrated into other wireless DL architectures\cite{10419367}, such as wireless generative artificial intelligence (GAI) and wireless LLMs. GNNs have been validated as augmenters for wireless GAI from both network and model perspectives, and for LLMs in terms of Graph-based Retrieval-Augmented Generation (RAG)\cite{edge_graphrag_0801}. In light of the growing number of GNN-based applications across layers\cite{10663726}, as well as the diversity of wireless technologies and scenarios, there is an urgent need to organize and implement GNNs in a more intelligent manner to align their capabilities with wireless task. 


Drawing inspiration from agentic AI, we put forward the concept of agentic GNNs. Unlike traditional GNNs, which passively infer outputs from inputs for single tasks, agentic GNNs enable multi-step inference by coordinating multiple GNNs, which is particularly suited for mobile communication systems with complex tasks. The agentic GNNs for enabling smart cities towards EGI are illustrated in Figure \ref{sys2}. GNNs  enhance feature extraction from wireless networks to adapt to time-varying channel conditions and network topologies, while agentic GNNs enable the adaptation of GNN capabilities (provided by different models) to diverse application tasks and scenarios, thereby achieving ubiquitous intelligent wireless coverage. Recently, there have emerged numerous GNNs tailored to diverse wireless scenarios and tasks. Exploring advanced applications of GNNs in the upcoming 6G remains a key aspect in constructing agentic GNNs. 

GNNs have spurred the development of various fundamental and practical communication designs and optimization schemes, including:
\begin{itemize}
    \item \emph{Real-time Processing \cite{li_twc_24_1,liang_jsac_2024_0612}:} GNNs can automatically learn to approximate the optimal mapping from network topologies to targeted signal processing strategies, routing results, and other outputs. The lightweight designs enable well-trained GNNs to exhibit high computational efficiency, thereby facilitating real-time computation. 

    \item \emph{Scalable and Distributed Implementation\cite{Zhang_twc_25,he_tmc_25}:} GNNs are scalable to elements in communication networks, enabling signal processing in ultra-dense and dynamic wireless environments. Such scalability  facilitates distributed implementation by integrating the signaling exchange of wireless networks with the message-passing mechanism of GNNs. 

    \item \emph{Unified Solution Framework\cite{han_tmc_25,songz_tvt_24}:} Learning from a graph representing a specific network system or topology can extract critical features. These features can be leveraged to tackle diverse problems through a unified framework. Fine-tuning pre-trained GNNs via transfer learning can further enhance the system's performance.

    \item \emph{Novel Problem Solver\cite{Singh_tgcn_24,tung_tvt_25}:} GNNs introduce a transformative paradigm for tackling complex communication problems from explicit mathematical formulation to data-driven learning. This paradigm shift allows GNNs to circumvent the computational intractability and oversimplified assumptions of suboptimal mathematical algorithms.



\end{itemize}

\subsection{Contributions}
Although there exist some surveys related to DL-driven wireless communications, most of them posit GNNs as plug-and-play solutions to network optimization, and overlook the fundamental mismatch between GNN architectures as well as learning strategies and wireless tasks. Table \ref{related} summarizes the most related surveys and their contributions. In particular, the early surveys \cite{sun_survey19,chen_survey19,zhang_survey19,Luong_survey19,Garadi_survey20,djigal_survey22,shi_survey23,zhou_survey24,du_survey24,khalek_survey24,boateng_survey25,zeng_survey25} explored  the ML and/or DL approaches (including DRL) for cross-layer designs and diverse scenarios, but overlooked GNNs as the core NN architectures. Recent surveys \cite{sun_survey19,chen_survey19,zhang_survey19,Luong_survey19,Garadi_survey20,djigal_survey22,shi_survey23,zhou_survey24,du_survey24,khalek_survey24,boateng_survey25,zeng_survey25} delved into task-oriented DL frameworks  for wireless networks and highlighted  the significance  of GNNs. However, they mainly discussed the vanilla message-passing GNNs and overlooked the graph representation methods of diverse wireless networks and the application of more advanced GNN architectures. Agentic GNNs necessitate more specific GNN applications for wireless communications and networking. To date, no survey has systematically addressed the application of GNNs to wireless networks. This gap motivates us to present this article, which combines an overview of GNN fundamentals and a comprehensive literature review of GNN applications in addressing wireless communication tasks. 

Notably, a full understanding of GNNs is crucial for developing agentic GNN frameworks. Specifically, we summarize the GNN structures for diverse wireless applications and emphasize the importance of aligning GNN architectures with wireless tasks. These include traditional network protocol stacks, emerging multifunctional wireless services, flexible antenna systems, and cell-free massive multiple-input multiple-output (CFmMIMO)\cite{7827017}. In addition to agentic GNNs for autonomous operation, we also propose the use of LLMs as question-and-answer (Q\&A) agents to assist operators or readers in quickly gaining insights into specific GNNs within agentic GNNs. The LLM framework is named SurveyLLM. For example, we use this survey and its references to construct a local knowledge base, and deploy an LLM locally to leverage this knowledge base to generate GNN-related responses tailored to wireless communication research. The key contributions of this article are summarized as follows:
\begin{itemize}
    \item \emph{Agentic GNN Framework.} We propose integrating existing and emerging GNN models into mobile communication systems towards EGI. These GNNs are orchestrated by agentic AI and adapt to diverse scenarios and tasks.
    
    \item \emph{Comprehensive Survey of Wireless GNNs.} We present a comprehensive overview of existing studies on GNN-enabled wireless designs with  focus on methods in graph representation and neural architectures, as well as the applications of GNNs in novel wireless techniques.

    \item \emph{Task-Oriented Comparative Analysis.} To gain in-depth insights into GNN-enabled wireless communications and networking, we systematically compare and analyze existing studies to highlight the alignment between GNN architectures and their applications.

    \item \emph{SurveyLLM as  Interactive Tools.} Given the growing literature on GNN models, we use this survey as a local knowledge base and LLMs as interactive tools to assist readers in learning about agentic GNNs in wireless communications and networking in a Q\&A fashion. The answer generation process is facilitated by the RAG framework.
    
    \item \emph{Open Problems and Future Research Directions.}   To further empower agentic GNNs, we highlight the necessity of studying graph representation for complex wireless networks, powerful GNN architectures for wireless networks, efficient schemes for handling complex constraints, model robustness maintenance, GNN-enabled wireless application layer, and guidelines for datasets, baselines, and evaluation.

\end{itemize}

\begin{table*}[t]
\centering
\footnotesize
\caption{An Overview of Selected Surveys in AI-enabled Wireless Communications}
\label{related}
\begin{tabular}{|c|c|c|c|c|c|c|c| }
\hline
\multirow{2}{*}{\bf Ref.}               & \multirow{2}{*}{\bf Year}    &  \multirow{2}{*}{\bf Contribution} &  \multicolumn{5}{c|}{\bf Focus of the Survey}        \\\cline{4-8}& &  &{\bf GNNs} &{\bf GR}$^{\dag}$ &{\bf NA}$^{\dag}$ & {\bf Agentic GNNs} & {\bf SurveyLLM} \\\hline\hline
\cite{sun_survey19}  & 2019 & Survey on ML for MAC, network, and application layers &$\times$ & \multirow{12}{*}{$\times$}  &  \multirow{12}{*}{$\times$} &  \multirow{12}{*}{$\times$} & \multirow{12}{*}{$\times$}\\\cline{1-4}
\cite{chen_survey19}   & 2019 & Survey on NNs-based  algorithms for wireless networking problems&  $\times$ &  & & &\\\cline{1-4}
\cite{zhang_survey19}   & 2019 & Survey on DL for mobile and wireless networking&  $\times$ & & &  &\\\cline{1-4}
\cite{Luong_survey19}   & 2019 & Survey on applications of DRL in communications and networking &  $\times$ & & &  &\\\cline{1-4}
\cite{Garadi_survey20}   & 2020 & Survey on ML and DL for IoT security &   $\times$ & & &  & \\\cline{1-4}
\cite{djigal_survey22}   & 2022 & Survey on  ML for resource Allocation in multi-access edge computing &  $\checkmark$ & & &  & \\\cline{1-4}
\cite{shi_survey23}   & 2023 & Survey on  ML for large-scale optimization  & $\checkmark$ & & &  &\\\cline{1-4}
\cite{zhou_survey24}   & 2024 & Survey on  ML-based optimization for
 RIS-Aided wireless networks  & $\checkmark$ & & & &\\\cline{1-4}
\cite{du_survey24}   & 2024 & Survey on generative diffusion models in network optimization & $\checkmark$ & & & &\\\cline{1-4}
\cite{khalek_survey24} & 2024 &Survey on ML-Driven cognitive radio for wireless networks  & $\checkmark$ & & & &\\\cline{1-4}
\cite{boateng_survey25}   & 2025 & Survey on  LLMs for communication, network, and service management  & $\times$ & & & &\\\cline{1-4}
\cite{zeng_survey25}   & 2025 & Survey on  empowering
 edge networks with graph intelligence  & $\checkmark$ & & & &\\\hline
\end{tabular}
\label{Overview of Selected Surveys}
\begin{tablenotes}
\footnotesize
\item $^{\dag}$GR and NA are short  for Graph Representation and  Neural Architecture, respectively.
\end{tablenotes}
\end{table*}






\begin{figure}[t]
\begin{center}
{\includegraphics[ width=1\linewidth]{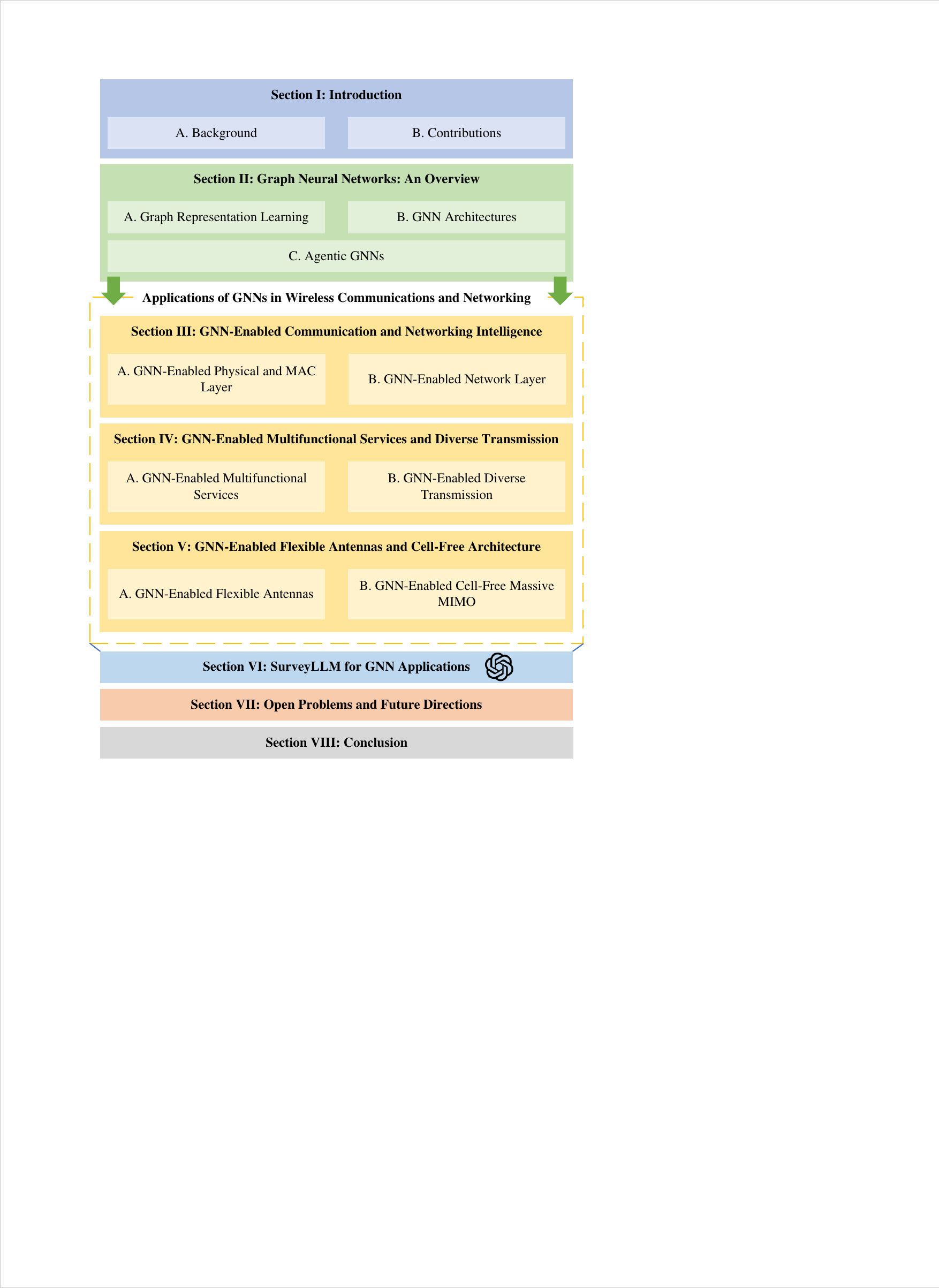}}
\caption{Outline of this article.}
\label{Outline}
\end{center}
\end{figure}

Survey Organization: We organize this article in a top-down hierarchy, as illustrated in Figure \ref{Outline}.  Section II provides an introduction to GNNs as well as the concept of agentic GNNs. Section III explores GNN-enabled communication and networking intelligence within conventional systems. Section IV reviews GNN applications for emerging multifunctional wireless services and diverse transmission scenarios.  Section V discusses GNN implementations in flexible-antenna systems and CFmMIMO. Section VI introduces an LLM framework built upon this article and the references.  Section VII  identifies critical challenges, open issues, and future directions. Finally, Section VIII concludes the article.

\begin{table*}
  \centering
  \footnotesize
  \caption{List of key technical abbreviations used throughout this survey.} 
  \label{abbreviations1}
  \begin{tabularx}{\textwidth}{lXlX}
    \toprule 
    \textbf{Abbreviation} & \textbf{Description} & \textbf{Abbreviation} & \textbf{Description} \\ 
    \midrule 
    AI	& artificial intelligence & AIGC & AI generated content
    \\
ANN & approximate nearest-neighbor & AO	& alternating optimization 
\\

BCE	 & binary cross-entropy  & BGNN	 & bipartite GNN \\
BP	 & belief propagation & BSUM 	 & block successive upper bound minimization \\
BiG-AMP & 	bilinear generalized approximate message passing 
& CNN 	 & convolutional NN  \\
ConvGNN	 &  convolutional GNN &
CoT	 & chain-of-thought \\
CRB& Cramér–Rao bound &
CSI & channel state information \\
DGAT	 & dynamic GAT & DGRL	 & deep GRL \\
DL 	 & deep learning &  SD3	 & soft deep deterministic policy gradient \\ DRL  & 	deep RL &
WMMSE & weighted minimum mean square error module \\
E-GNN	 & enhanced-GNN &
 EGAT &  edge-enhanced GAT \\
 EGNN & equivariant GNN & EGRN & edge aggregated graph attention regression network \\
 FANET & 	flying ad-hoc network &
 GAI &  generative artificial intelligence\\
GAT & graph attention network&
GCN & graph convolutional network \\
GELU & Gaussian error linear unit &
GINEConv & 	graph isomorphism network with edge features convolution \\ GNN 	 & graph NN &
GraphConv & graph convolution \\ 
GRU & gated recurrent unit &
 GraphSAGE	 &  graph sample and aggregate \\
 HAN & heterogeneous attention network&
 HetGNN & heterogeneous GNN \\
HGNN & Heterogeneous GNN & KL & 	Kullback-Leibler 
\\
 LSTM  & 	Long Short-Term Memory 
&
LLM		 &  large language model \\
 LMMSE & 	linear minimum mean square error 
&
MDP & 	Markov decision process\\
ML 	 & 	machine learning & MLP	 & multi-layer perceptron 
\\
  MPNN	 &  message passing NN &
 MSE	 & mean square error \\  NMSE	 & normalized MSE 
&
NN 		 &  neural network \\ OTA		 & over-the-air 
&
PathGNN	 & path-based GNN \\
PPO	 &  proximal  policy optimization 
&
RAG	 & retrieval-augmented generation \\ RecGNN  & 	recurrent GNN 
 &  RL 	 & 	reinforcement learning
\\
 RNN 	 & 	recurrent neural network &
SCA 	 & 	successive convex approximation
\\
SDR & 	semidefinite relaxation &  SOCP & 	second-order cone programming
\\
SVM 	 & support vector machines & 
TE	 & traffic engineering 
\\
    \bottomrule
  \end{tabularx}
\end{table*}

\section{Graph Neural Networks: An Overview}

This section outlines the preliminaries and backgrounds of graph representation learning and GNN architectures. Furthermore, we elaborate on the advantages of agentic GNNs. For clarity, we present a unified notational convention: boldface uppercase letters denote matrices, and boldface lowercase letters denote vectors. All symbols used in this paper are  defined in Table \ref{notation}. 



\subsection{Graph Representation Learning}


This subsection introduces the fundamental concepts of graphs and graph representation learning \cite{GRL}. 

\subsubsection{Graph} A graph is a non-Euclidean data structure consisting of nodes and the interactions between node pairs. For example, in social networks, individuals are represented as nodes and friendships as edges, facilitating the modeling of interpersonal relationships and community detection. Moreover, communication systems involve many graph-structured topologies. For instance, in mobile systems, BSs and users act as nodes with links as edges; in core networks, routers serve as nodes while traffic flows are represented as edges.

Mathematically, a graph is represented as $\mathcal{G} = (\mathcal{V}, \mathcal{E})$, where $\mathcal{V}$ denotes the set of nodes (or vertices), and $\mathcal{E}$ denotes the set of edges. Specifically, $v_i \in \mathcal{V}$ represents a node, and $e_{ij} = \langle v_i, v_j\rangle \in \mathcal{E}$ represents a  directed edge from $v_j$ to $v_j$. The neighborhood of a node $v_i$ is defined as $\mathcal{N}(v_i) = \{u_j \in \mathcal{V} \mid e_{ij} \in \mathcal{E}\}$, and the adjacency matrix $\mathbf{A}$ is defined such that ${\bf A}_{ij} = 1$ if $e_{ij} \in \mathcal{E}$ and ${\bf A}_{ij} = 0$  if $e_{ij} \notin \mathcal{E}$. For graphs modeling multiple interactions, they comprise different types of edges, where $\cal R$ denotes the set of relations. Such graphs are often referred to as heterogeneous and multiplex graphs. In  the former, nodes can also be imbued with types, where certain edges typically connect nodes of specific types; in the latter, the graph is decomposed into several layers  (each representing a unique relation), where nodes belong to all layers, and both inter-layer and intra-layer edges exist to connect nodes within the same layer and across different layers. Lastly, a graph is associated with feature or attribute information. Such features are represented by real- or complex-valued matrices, which are assigned to nodes, edges, or even the graph.

\subsubsection{Graph representation learning} Graph representation learning aims to enable ML on graph-structured data, with the goal of learning representations that encode the structural information of the graph to enhance downstream tasks. These tasks include node classification, relation prediction\cite{2023arXiv230700865L}, clustering and community detection\cite{10106414}, as well as graph classification, regression, and clustering\cite{tsitsulin2023graph}. Typically, graph representation learning methods can be viewed as an encoder-decoder framework, where the encoder maps nodes{\footnote{Here, we take node-centric graphs as an example, where only node features are considered. This definition is also applicable to edge-centric graphs.}} to embeddings, and the decoder uses these embeddings to produce desired results\cite{10518175}. The embeddings encapsulate both the nodes' positions in the graph and the structure of their local neighborhoods. This framework has given rise to several well-known graph representation learning methods, including Laplacian eigenmaps\cite{shuman2013emerging}, graph factorization\cite{10.1145/2488388.2488393}, GraRep\cite{cao2015grarep}, HOPE\cite{8329541}, DeepWalk, and node2vec\cite{10.1145/2939672.2939754}, which differ in their decoders and loss functions.

DL has been applied to advance graph representation learning. However, CNNs for grid-structured input and recurrent NNs (RNNs)\cite{fang2021survey} for sequential data are often unsuitable for graph-structured data. GNNs have been proposed  as a specialized DL framework for processing graph-structured data to tackle graph representation learning problems, capable of modeling complex interactions among nodes. In general, GNNs are required to satisfy either permutation invariance or permutation equivariance\cite{qu2023concept}. Mathematically, a function $f$ (which can be constructed using GNNs) being permutation invariant or equivariant is expressed as   
\begin{flalign}
    &f\left({\bf P}{\bf A}{\bf P}^T\right) = f\left({\bf A}\right),~{\rm (Permutation~Invariance)},\\
    &f\left({\bf P}{\bf A}{\bf P}^T\right) = {\bf P}f\left({\bf A}\right),~{\rm (Permutation~Equivariance)},
\end{flalign}
where $\bf P$ denotes a permutation matrix. Specifically, permutation invariance means that GNNs are independent of the arbitrary ordering of rows and columns in the adjacency matrix, while permutation equivariance implies that the output of GNNs undergoes a consistent permutation when the adjacency matrix is permuted. Permutation equivariance is crucial in wireless networks, as it enhances generalization ability and reduces the sample complexity and training costs \cite{Shen_jsac_21}.

\begin{table}[t]
\centering
\footnotesize
\caption{Descriptions of Notations and Functions used in Graph Neural Networks.}\label{notation}
\begin{tabular}{|l|l|}
\hline
\textbf{Symbol} & \textbf{Definition} \\
\hline\hline
$\cal G= ({\cal V}, {\cal E})$ & A graph with node set  $\cal V$ and edge set $\cal E$. \\
$v_i$, $v_j\in {\cal V}$ & A node in the graph. \\
$e_{ij}\in {\cal E}$ & An edge in the graph. \\
${\cal N}(v_i)$ & The neighbors of node $v_i$. \\
$\mathbf{A}\in\mathbb{R}^{|\mathcal{V}|\times|\mathcal{V}|}$ & The adjacency matrix. \\
$\mathbf{D}$ & The degree matrix with $\mathbf{D}_{ii} = \sum_{j=1}^{n} \mathbf{A}_{ij}$. \\
$\mathbf{x}_{v_i}$ & The feature vector of  node $v_i$. \\
$\mathbf{x}_{e_{ij}}$ & The feature vector of  node $e_{ij}$. \\
$\mathbf{h}_{v_{i}}$, ${\bf h}_{e_{ij}}$ & The  hidden features of node and edge. \\
$^{(l)}$, $^{(l+1)}$ & Superscript representing the layer index. \\
$\mathbf{W}$, $\mathbf{W}_e$ & Learnable parameters for  transformation. \\
$\bf a$ & Learnable parameter for attention. \\
$a_{ij}$ & Attention coefficient between $v_i$ and $v_j$.\\
\hline
\hline
\textbf{Function} & \textbf{Definition} \\
\hline
$\sigma(\cdot)$ & Sigmoid activation function. \\
${\rm exp}(\cdot)$  & Exponential function.\\
$\|$ & Concatenation operation\\
${\rm LeakyReLU}(\cdot)$  & Leaky rectified linear unit.\\
${\rm Softmax}(\cdot)$  & Soft maximum function.\\
${\rm tanh}(\cdot)$  & Hyperbolic tangent 
function.\\
${\rm GRU}(\cdot)$ & Gated Recurrent Unit cell function.\\
\hline
\end{tabular}
\end{table}

\subsection{GNN Architectures}

\begin{figure*}[t]
    \centering
    \includegraphics[width=0.9\textwidth]{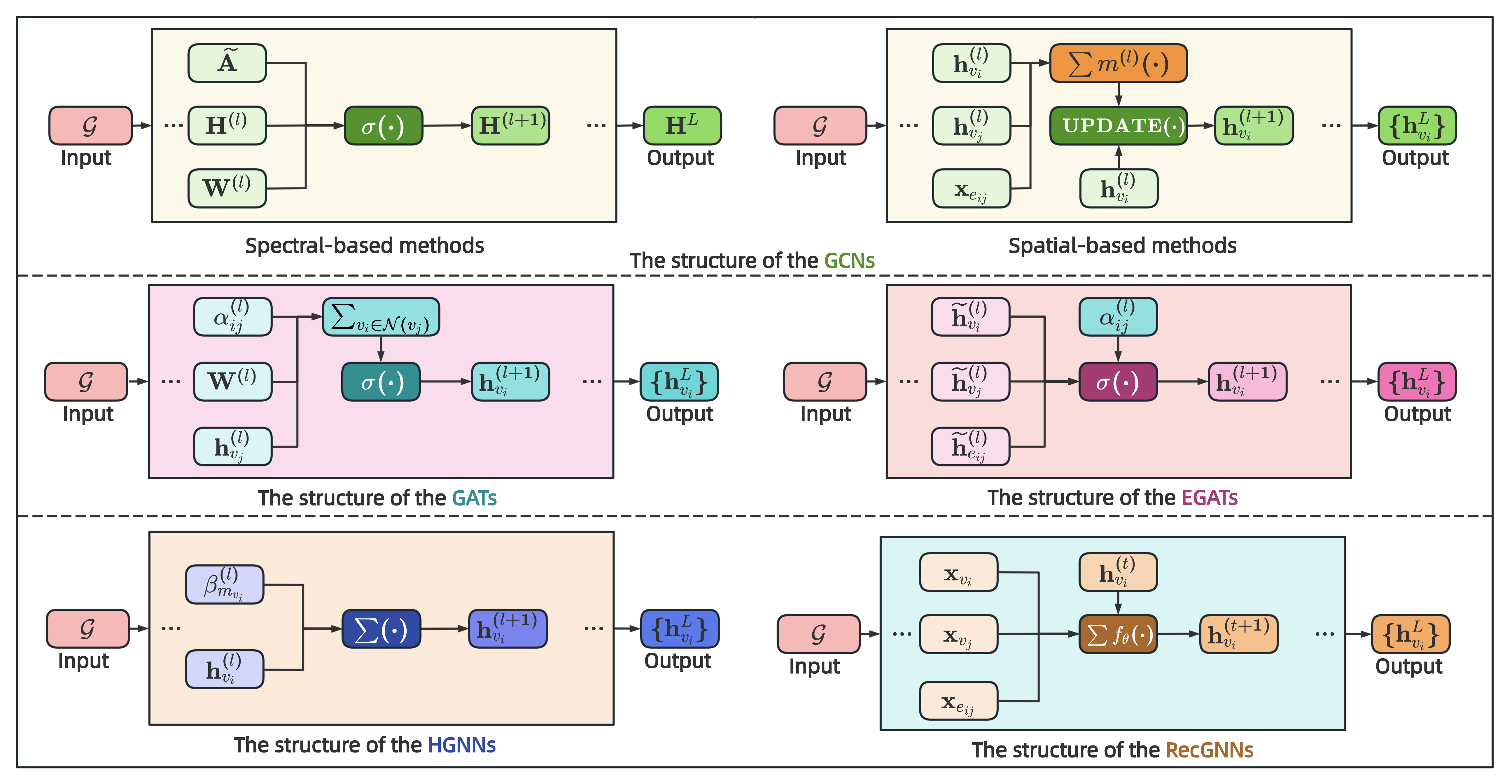}
    \caption{The structural schematics of five typical GNN architectures: GCNs are divided into spectral-based and spatial-based structures, performing feature aggregation and updates; GATs and EGATs highlight the role of attention coefficients in feature aggregation, with EGATs further incorporating edge feature processing; HGNNs illustrate feature processing and aggregation methods tailored for heterogeneous graphs; RecGNNs demonstrate a recurrent update mechanism, achieving node state updates through iteration over time steps.}
    \label{fig:GGNN_GCN}
\end{figure*}

\begin{table*}[t]
\centering
\footnotesize
\caption{Summary of common GNN models.}
\label{GNNs}
\begin{tabular}{|c|
>{\raggedright\arraybackslash}m{4.82cm}|
>{\raggedright\arraybackslash}m{4.82cm}|
>{\raggedright\arraybackslash}m{4.82cm}|
>{\raggedright\arraybackslash}m{4.82cm}| }
\hline
{\bf Model} & {\bf Key features}  & {\bf Pros} & {\bf Cons} \\ \hline\hline

\multirow{4}{*}{GCN \cite{bruna2013spectral, zhang2019bayesian}} &
Spatial-based: aggregates neighbor features via message-passing &
Spatial-based: Scalable to large/dynamic graphs, supports inductive learning &
Spectral-based: Eigen-decomposition, poor scalability and transferability \\ 
\cline{2-4}
& Spatial-based: aggregates neighbor features via message-passing &
Spatial-based: Scalable to large/dynamic graphs, supports inductive learning &
Over-smoothing, lacks spectral interpretability \\ \hline

GAT \cite{Velickovic2017GraphAN} &
Introduces an attention mechanism to weigh neighbors &
Adaptive to local structure, supports heterogeneous graphs &
Computationally expensive (multi-head), indistinct weights in deep layers \\ \hline

EGAT \cite{wang2021egat} & 
Enhances GAT by incorporating edge features into attention & 
Encodes edge semantics, suitable for edge-attributed graphs & 
More complex and costly than GAT for high-dimensional edge features \\ \hline
HGNN \cite{Wang_HAN_2019} & 
Uses metapaths/metagraphs to model heterogeneous structure & 
Captures semantic relations in heterogeneous graphs & 
Relies on expert knowledge, complex to train \\ \hline
RecGNN \cite{Li_GGSNN_2017} & 
Applies RNN/GRU recurrent updates to capture state transitions & 
Retains state memory, suitable for dynamic/sequential graphs & 
Inefficient on large graphs due to iteration and memory \\ \hline
\end{tabular}
\end{table*}


Most existing GNNs are built upon message-passing iterations. One message-passing iteration involves two key processes: ${\rm AGG}(\cdot)$ denoting aggregation and ${\rm UPDATE}(\cdot)$ representing updating\cite{sunil2012message}, mathematically expressed as
\begin{flalign}
    &{\bf h}^{\left(l+1\right)}_{v_i}=\\
    &{\rm UPDATE}\left({\bf h}^{\left(l\right)}_{v_i}, \underbrace{{\rm AGG}^{\left(l\right)}\left(\left\{{\bf h}^{\left(l\right)}_{v_j},\forall {u_j}\in{\cal N}\left({v_i}\right)  \right\}\right)}_{\triangleq {\bf m}^{\left(l\right)}_{{\cal N}\left({v_i}\right)}} \right)\nonumber.
\end{flalign}
Here, ${\bf m}_{{\cal N}\left({v_i}\right)}^{\left(l\right)}$ denotes the ``message" that aggregates the features of the target node's neighborhood. The message-passing mechanism enables nodes to capture interactions with other nodes as well as global structural information. The message passing gives rise to the message-passing NNs (MPNNs) \cite{10.5555/3305381.3305512}, which serve as a unified framework for GNNs. On the other hand,  some GNNs eliminate the iterative message-passing process. In the following, we present four message-passing-based GNNs, i.e., Graph Convolutional Network (GCN), Graph Attention Network (GAT), Edge-enhanced GAT (EGAT), and Heterogeneous GNN (HGNN), as well as a GNN not inherently based on message passing, i.e., Recurrent GNN (RecGNN). The neural architectures of these models are illustrated in Figure \ref{fig:GGNN_GCN} with their pros and cons summarized in Table \ref{GNNs}.


\subsubsection{GCN}

GCNs constitute a fundamental class of GNNs. Based on how convolution operations are defined on graphs, GCNs are categorized into two types: spectral-based  and spatial-based convolutional GNNs (ConvGNNs).

Spectral-based ConvGNNs originate from graph signal processing\cite{Sandryhaila_ITSP_130110}, where graph convolution (GraphConv) is defined in the frequency domain. Define the symmetric normalized Laplacian matrix as:
\begin{flalign}
{\bf L} = {\bf I}_{|\cal V|} - {\bf D}^{-\frac{1}{2}} {\bf A} {\bf D}^{-\frac{1}{2}}.
\end{flalign}
By performing eigendecomposition $\mathbf{L} = \mathbf{U} {\bm\Lambda} \mathbf{U}^T$, we obtain the spectral convolution of a node feature signal as
\begin{flalign}\label{sc}
\mathbf{x}_{v_i}  \star_{\cal G} \mathbf{g}  = {\bf U}\left(\left({\bf U}^T{\bf x}_{v_i} \right)\odot \left({\bf U}^T\mathbf{g} \right)\right),
\end{flalign}
where $\star_{\cal G}$ denotes the convolution specific to $\cal G$ and $\mathbf{g}$ represents a spectral filter\cite{bruna2013spectral}. To improve computational efficiency, ChebNet\cite{Defferrard_Anips_2016} approximates \eqref{sc} using Chebyshev polynomials, simplifies it to a first-order approximation, and results in the following layer-wise propagation rule to update node features as
\begin{flalign}
\mathbf{H}^{(l+1)} = \sigma\left(\tilde{\mathbf{A}} \mathbf{H}^{(l)} \mathbf{W}^{(l)}\right),
\end{flalign}
where $\tilde{\mathbf{A}} = ({\bf D}+\mathbf{I}_{|\cal V|})^{1/2}(\mathbf{A} + \mathbf{I}_{|\cal V|})({\bf D}+\mathbf{I}_{|\cal V|})^{1/2}$ is a normalized variant of the adjacency matrix (with self-loops).

The advantages of spectral-based ConvGNNs lie in their rigorous theoretical foundation and global receptive field. However, their drawbacks include reliance on eigen-decomposition, poor scalability, and limited transferability.

Spatial-based ConvGNNs define convolutions directly in the graph domain via the message-passing framework. They model the propagation of information between nodes by aggregating features from each node’s local neighborhood. MPNNs provide a general framework\cite{zhang2019bayesian} for spatial-based ConvGNNs by abstracting GraphConv as a message-passing process, with the update rule given by
\begin{flalign}
{\bf h}_{v_{i}}^{\left(l+1\right)} = {\rm UPDATE}\left({\bf h}_{v_i}^{\left(l\right)}, \sum_{v_j \in {\cal N}(v_i)} m^{\left(l\right)}\left({\bf h}_{v_{i}}^{\left(l\right)}, {\bf h}_{v_{j}}^{\left(l\right)}, {\bf x}_{{e_{ij}}}\right)\right),
\end{flalign}
where $ m^{\left(l\right)}(\cdot)$ denotes the message function, and the summation operation aggregates messages from all neighbors of node $v_i$. Different models adopt distinct aggregation strategies. For instance, Graph Sample and Aggregate (GraphSAGE) employs mean, Long Short-Term Memory (LSTM), or pooling aggregators \cite{10.5555/3294771.3294869}.

Spatial-based ConvGNNs offer advantages including scalability to large and dynamic graphs\cite{pareja2020evolvegcn}, flexibility, and inductive capability. Their drawbacks, however, include a tendency toward over-smoothing issues \cite{LI_tpami_23} and a lack of rigorous spectral foundations.

\subsubsection{GAT}

GATs\cite{Velickovic2017GraphAN} incorporate an attention-based mechanism into the message-passing framework, enabling the model to assign varying importance weights to neighbors during feature aggregation. Unlike GCNs, which utilize fixed and degree-based normalization coefficients, GATs learn adaptive attention weights in a data-driven manner.

The attention coefficient $\alpha_{ij}^{(l)}$ assigned by node $v_i$ to node $v_j$ is given by
\begin{flalign}\label{self-att}
\alpha_{ij}^{(l)} = \frac{\exp\left(a_{ij}^{(l)}\right)}{\sum_{v_j\in {\cal N}(v_i)} \exp\left(a_{ij}^{(l)}\right)},
\end{flalign}
where
\begin{flalign}
a_{ij}^{(l)} = \text{LeakyReLU}\left( \left(\mathbf{a}^{(l)}\right)^T \left[ \mathbf{W}^{(l)} \mathbf{h}_{v_i}^{(l)} \,\|\, \mathbf{W}^{(l)} \mathbf{h}_{v_j}^{(l)} \right] \right).
\end{flalign}

Then, node $v_i$ updates its feature by
\begin{flalign}
\mathbf{h}_{v_i}^{(l+1)} = \sigma\left( \sum\nolimits_{v_i \in {\cal N}\left(v_j\right)} \alpha_{ij}^{(l)} \mathbf{W}^{(l)} \mathbf{h}_{v_j}^{(l)} \right).
\end{flalign}

GATs can be further extended by incorporating the multi-head attention mechanism\cite{cheng2021drug}, which employs multiple independent attention heads, as follows:
\begin{flalign}
\mathbf{h}_{v_i}^{(l+1)} = \mathbin\Vert_{d=1}^D \sigma \left( \sum\nolimits_{v_j \in {\cal N}\left(v_i\right)} \alpha_{d,ij}^{(l)} \mathbf{W}^{(l)}_d \mathbf{h}_{v_j}^{(l)} \right),
\end{flalign}
where $D$ is the number of attention heads; and $\alpha_{d,ij}^{(l)}$  and $ \mathbf{W}^{(l)}_d$ are the attention coefficient and learnable parameters associated with the $d$-th head, respectively.

GATs are capable of capturing local structural variations within the graph by assigning adaptive and context-aware weights to neighboring nodes. Moreover, GATs support inductive learning\cite{xiao2023adversarially} and function independently of the graph Laplacian, thereby enhancing their flexibility across diverse scenarios. However, GATs also have limitations, such as computational complexity linear to the number of attention heads and graph size, and indistinguishable attention weights in deep networks (cf. the over-smoothing issue).  

\subsubsection{EGAT}

EGATs\cite{wang2021egat} extend the canonical GATs by integrating edge features into the attention mechanism. They are able to capture richer relational semantics, a capability that is particularly critical in edge-rich graphs such as knowledge graphs, transportation graphs \cite{zeng2024estimating}, molecular graphs, and ultra-dense communication networks.


The features pertaining to a pair of nodes and their connecting edge are first linearly transformed into 
\begin{flalign}
{\widetilde {\bf h}}_{v_i}^{(l)} = \mathbf{W}^{(l)} \mathbf{h}^{(l)}_{v_i}, ~ 
{\widetilde {\mathbf h}}_{v_j}^{(l)} = \mathbf{W}^{(l)} \mathbf{h}^{(l)}_{v_j}, ~ 
{\widetilde {\mathbf h}}_{e_{ij}}^{(l)} = \mathbf{W}_e^{(l)} {\bf h}^{(l)}_{{e_{ij}}}.
\end{flalign}

Next, the attention coefficient $\alpha_{ij}^{(l)}$ between a target node $v_i$ and its neighbor $v_j \in {\cal N}(v_i)$ is computed via a shared attention mechanism that jointly incorporates node features and edge features:
\begin{flalign}
&\alpha_{ij}^{(l)} = \\
&{\rm Softmax}{
\left( \mathrm{LeakyReLU}\left( \left(\mathbf{a}^{(l)}\right)^T \left[ {\widetilde {\mathbf h}_{v_i}^{(l)}} \,\|\, {\widetilde{\mathbf h}}_{v_j}^{(l)} \,\|\, {\widetilde {\mathbf h}}_{e_{ij}}^{(l)} \right] \right) \right).
}\nonumber
\end{flalign}

Node feature aggregation is then performed via attention-weighted summation of neighbor features:
\begin{flalign}
\mathbf{h}_{v_i}^{(l+1)} = \sigma \left( \sum\nolimits_{v_j \in {\cal N}\left(v_i\right)} \alpha_{ij}^{(l)} \mathbf{W}^{(l)} \mathbf{h}^{(l)}_{v_i} \right).
\end{flalign}
Similarly, the edge feature is updated  by
\begin{flalign}
\mathbf{h}_{e_{ij}}^{(l+1)} = \sigma \left( \sum\nolimits_{{e_{i^{\prime}j^{\prime}}} \in {\cal N}\left({e_{ij}}\right)} \mu_{e_{ij},{e_{i^{\prime}j^{\prime}}}}^{(l)} \mathbf{W}_e^{(l)} {\bf h}^{(l)}_{{e_{ij}}} \right),
\end{flalign}
where ${\cal N}\left({e_{ij}}\right)$ denotes the first-order neighbor set of $e_{ij}$, and $\mu_{e_{ij},{e_{i^{\prime}j^{\prime}}}}^{(l)}$ denotes the attention weight assigned by $e_{ij}$ to ${e_{i^{\prime}j^{\prime}}}$, which is computed via the attention mechanism.  

Compared to standard GATs, EGATs explicitly encode edge semantics, enabling more informed and context-aware attention computations. Such edge awareness enhances the model’s discriminative power, making it well-suited for relational, heterogeneous, or edge-attributed graphs. Moreover, EGATs demonstrate strong generality, allowing them to be effectively applied to a broad range of graph-based learning tasks.

\subsubsection{HGNN}

GNNs discussed above are fundamentally designed for homogeneous graphs, where all nodes and edges  belong to a single type. However, real-world graphs often exhibit rich heterogeneity, resulting in diverse node and edge types. Graph heterogeneity can render conventional GNNs tailored for homogeneous graphs ineffective. To model the intricate relationships between nodes,  the \emph{metapath} and \emph{metagraph} structures have been proposed. Specifically, a metapath represents a sequence of node types connected by edges to capture semantic specificity, while a metagraph represents a structured aggregation or extension of multiple metapaths to convey finer-grained semantics. Then, one effective approach to addressing graph heterogeneity is to comprehensively fuse features of different node and edge types.

Leveraging attention mechanisms, Heterogeneous Attention Network (HAN)\cite{Wang_HAN_2019} can learn node embeddings following metapaths in an end-to-end manner. It is mainly composed of three key components: node-level attention, semantic-level attention, and feature updating. First, HAN computes the importance of neighbors using a self-attention mechanism (cf. \eqref{self-att}) for each metapath, and aggregates the weighted representations for each node. Then, the semantic-level attention fuses semantic-specific features from multiple metapaths, and each metapath is assigned an attention weight given by
\begin{flalign}
\beta_{m_{v_i}}^{(l)} = \frac{\exp\left(w^{(l)}_{m_{v_i}}\right)}{\sum\nolimits_{{{m^{\prime}_{v_i}}\in{\cal M}\left(v_i\right)}} \exp\left(w^{(l)}_{{m^{\prime}_{v_i}}}\right)},
\end{flalign}
with
\begin{flalign}
w_{m} = \frac{1}{|\mathcal{V}|} \sum_{{v_i} \in \mathcal{V}} \left(\mathbf{q}^{(l)}\right)^T \tanh\left( \mathbf{W}_m^{(l)} \mathbf{h}_{m,{v_i}}^{(l)} + \mathbf{b}_m^{(l)} \right),
\end{flalign}
where ${\cal M}(v_i)$ denotes the set of metapaths associated with node $v_i$, $\mathbf{h}_{m,{v_i}}^{(l)}$ represents the metapath-aware node features, and $\mathbf{q}^{(l)}$, $\mathbf{W}_m^{(l)}$, and $\mathbf{b}_m^{(l)}$ are the corresponding learnable parameters. Finally, the node feature in the prediction layer is computed by
\begin{flalign}
\mathbf{h}_{v_i}^{(l+1)} = \sum\nolimits_{m_{v_i}\in{\cal M}\left(v_i\right) } \beta^{(l)}_{m_{v_i}}  \mathbf{h}_{v_i}^{(l)}.
\end{flalign}
A task-specific layer is applied to the learned node features to facilitate downstream tasks such as node classification, clustering, and link prediction.

\subsubsection{RecGNN}

RecGNNs employ the gating methods from RNNs to improve learning ability and stability. They extract high-level node representations by recurrently applying the same set of parameters across nodes in a graph. The earliest RecGNN models were  limited by computational power and restricted to handling directed acyclic graphs (DAGs)\cite{Sperduti_ITNN_0531,Micheli_ITNN_041108}. RecGNNs leverage an information diffusion mechanism, where node states are updated through repeated information exchange with neighbors until a global equilibrium. A node’s hidden state is recurrently updated by
\begin{flalign}
{\bf h}_{v_i}^{(t+1)} = \sum\nolimits_{{v_j} \in \mathcal{N}(v_i)} f_{\bm\theta}\left({\bf x}_{v_i}, {\bf x}_{v_j},{\bf x}_{e_{ij}},{\bf h}^{(t)}_{v_i}\right),
\end{flalign}
where $t$ and $(t+1)$ denote the time step, and the recurrent function $f_{\bm\theta}(\cdot)$ is parameterized by ${\bm\theta}$.

Later, gated recurrent units (GRUs) are incorporated to replace traditional recurrent functions, eliminating the need to constrain the parameters to ensure convergence. The Gated GNNs (GGNNs) \cite{Li_GGSNN_2017} further restrict recurrent update steps to a fixed number, thereby eliminating the need for convergence.
\begin{flalign}
{\bf h}_{v_i}^{(t+1)} = \mathrm{GRU}\left({\bf h}_{v_i}^{(t)}, {\bf m}_{{\cal N}\left({v_i}\right)}^{\left(t\right)}\right).
\end{flalign}
However, such recurrent models are unsuitable for large-scale graphs, as they involve multiple iterations and necessitate storing the intermediate states of all nodes at each iteration step, leading to high memory overhead. To address the limitation, Stochastic Steady-State Embedding (SSE) was proposed for periodic updating of node hidden states in a stochastic and asynchronous manner \cite{dai2018learning}.

\subsection{Agentic GNNs}
Agentic AI aims to enable sophisticated reasoning and iterative planning to autonomously handle complex tasks under limited supervision. We adapt agentic AI, originally designed for general applications, to agentic GNNs tailored to wireless systems, because most network layers employ GNNs as fundamental models to enhance their service capabilities. Unlike traditional GNNs designed for single tasks in isolation, agentic GNNs are characterized by enabling multi-step problem-solving\cite{sapkota2025aiagentsvsagentic}. For instance, in  ISAC applications, agentic GNNs can decompose a task into a series of subtasks: GNN-enabled channel estimation, GNN-enabled joint information and sensing beamforming, GNN-enabled channel decoding and signal detection, and GNN-enabled sensing information reconstruction and generation. Each subtask is handled by a task-specific GNN, and all GNNs operate autonomously and coordinately to achieve the goal, i.e., providing high-quality ISAC services to mobile users.
 
\begin{figure}[t]
\begin{center}
{\includegraphics[ width=1\linewidth]{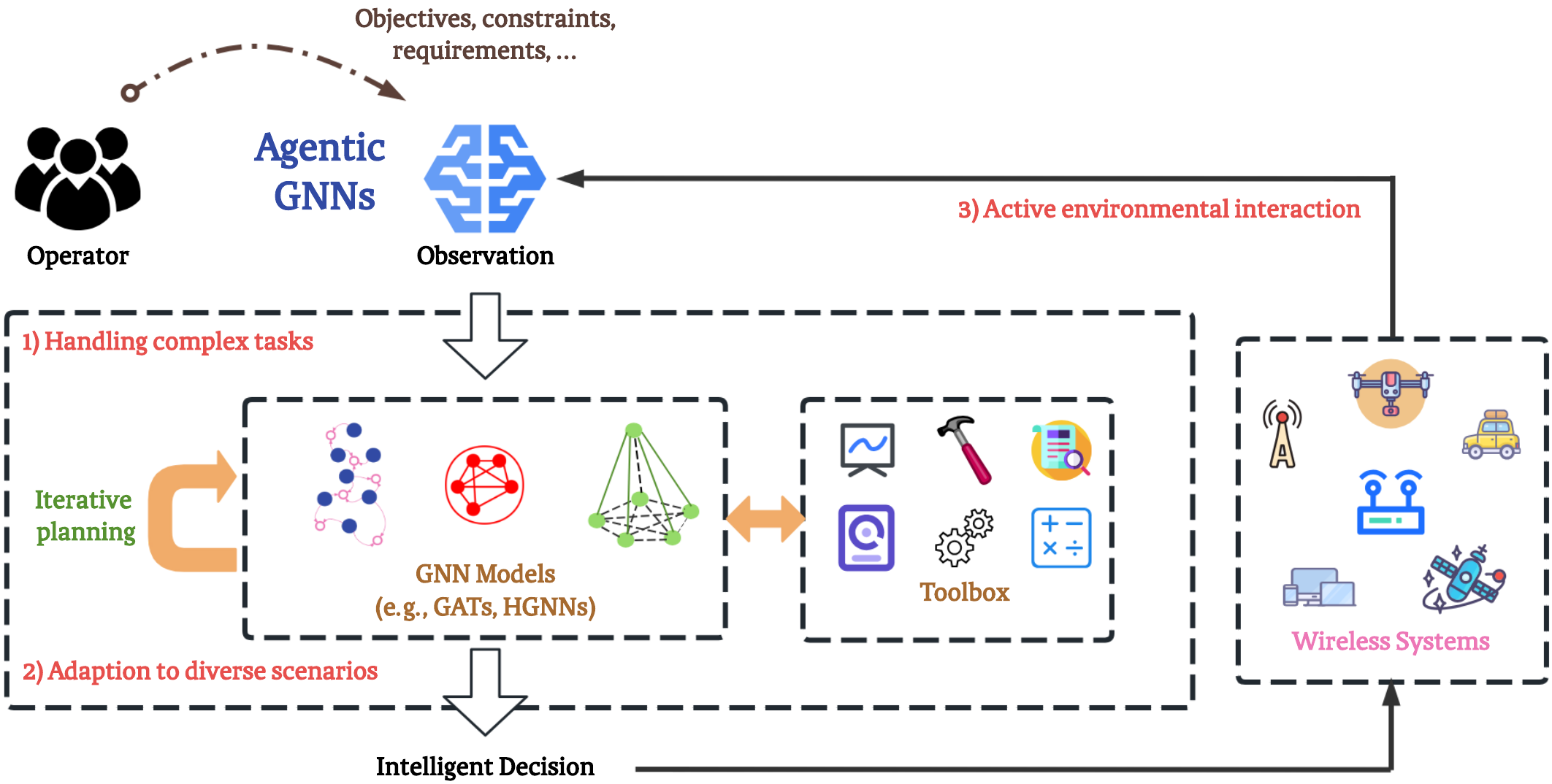}}
\caption{Illustration of Agentic GNNs.}
\label{AgenticGNN}
\end{center}
\end{figure}

The framework of agentic GNNs is illustrated in Figure \ref{AgenticGNN}. Operators can set objectives, constraints, and requirements for agentic GNNs; subsequently, agentic GNNs continuously collect observations from wireless systems, feed them into multiple GNNs, perform iterative planning with the assistance of toolboxes, and generate intelligent decisions.  Compared to traditional AI, agentic GNNs offer significant advantages on the following respects.

\begin{itemize}
    \item \emph{Handling complex tasks.} In mobile communication systems, mobile applications may require diverse services, including communication, sensing, navigation, and computation, from their serving BSs, resulting in multiple multi-step tasks. These systems need to continuously optimize resources to meet time-varying requirements. Agentic GNNs can perform iterative planning with multiple GNN models to maintain high performance.
    \item \emph{Adaptation to diverse scenarios.} Mobile communication systems shall be equipped with multiple novel transmission techniques, and these techniques are powered by diverse GNN models due to inherent hardware differences (cf. Section V). Additionally, mobile communication systems can be implemented in terrestrial, aerial, and satellite modes, resulting in numerous application scenarios. Agentic GNNs can match appropriate GNNs to specific scenarios and coordinate multiple GNNs to achieve an integration gain.
    \item \emph{Active environmental interaction.} Wireless environments are inherently dynamic, characterized by time-varying channel conditions, heterogeneous deployment scenarios, and strict quality of service (QoS) constraints. This challenges the availability, consistency, and comprehensiveness of data, leading to unstable performance of GNNs. Agentic GNNs actively collect and store experiences across task cycles via environmental interaction, further improving data-driven GNNs.
\end{itemize}

The above characteristics enable agentic GNNs to deploy GNNs to facilitate wireless communications and networking in an autonomous manner. This allows researchers to focus on developing more GNNs and integrating their capabilities into agentic GNNs.



\subsection{Summary}

In this section, we present a fundamental overview of graph representation learning and GNNs primarily based on message passing. We discuss the alignment between GNNs and wireless networks, focusing on graph representations of network topologies and the GNN layers tailored to wireless signal processing requirements. We also present the fundamental principles of applying GNNs to signal processing. Finally, we elaborate on the necessity and advantages of agentic GNNs, thereby enabling us to focus on the capabilities of existing GNNs. In subsequent sections, we survey state-of-the-art GNN applications for optimizing wireless communication systems and network protocols. 

\section{{GNN-Enabled Communication and Networking Intelligence}}

\begin{table*}[t]
\centering
\footnotesize
\caption{GNN-Enabled Communication and Networking Intelligence.}
\begin{tabular}{|c||c|c|c|c|c|c|c|c| }
\hline
{\bf Layer} &  {\bf Ref.}&{\bf Graph} &{\bf Task} & {\bf Problem}  & {\bf Model}  &   {\bf Train}  &{\bf Metrics} &  {\bf Scalability}        \\ \hline\hline
\multirow{24}{*}{\thead{Physical\\ \& MAC}} & \cite{li_twc_24_1} &\thead{Homogeneous,\\fully-connected,\\undirected}& \thead{Beamforming\\design 
 } & \thead{QoS-constrained\\sum-rate  max.}  & GAT &    \thead{Unsupervised,\\ penalty terms,\\Lagrangian,\\
duality} & \thead{Optimality,\\ scalability,\\ feasibility rate,\\inference time} & UE  \\    
 \cline{2-9}
  & \cite{liang_jsac_2024_0612} & \thead{Bipartite,\\
undirected}& \thead{Robust\\beamforming\\design 
 } & \thead{Outage-constrained\\sum-rate  max.\\\& power min.}  & BGNN & \thead{Unsupervised,\\model-driven,\\ penalty terms} & \thead{Optimality,\\ scalability,\\ feasibility rate,\\inference time} & \thead{Antenna, UE}  \\    
 \cline{2-9}
  & \cite{hu_TVT_24_0718} & \thead{Homogeneous,\\directed} & \thead{Power allocation} & \thead{Sum-rate max.} & \thead{UPAR} & \thead{Unsupervised} & \thead{Optimality,\\inference time} & \thead{N/A}  \\ 
 \cline{2-9}
  & \cite{Deng_jsac_25} & \thead{Homogeneous,\\fully-connected,\\directed} & \thead{Hybrid\\beamforming\\design} & \thead{Sum rate max.} & \thead{SA-MUHBF} & \thead{Unsupervised,\\model-driven} & \thead{Optimality,\\ scalability} & \thead{Antenna, UE}  \\ 
  \cline{2-9}
  & \cite{Liu_twc_25} &\thead{Homogeneous,\\fully-connected,\\directed}& \thead{MIMO data \\detection  \\and channel\\
estimation} & \thead{JCD} & \thead{GNN-\\BiGAMP,\\BiGNN-\\BiGAMP} & \thead{Supervised,\\pre-training,\\multi-task} & \thead{NMSE, SER} & \thead{N/A}  \\ 
  \cline{2-9}
  & \cite{Xuan_tnse_22} & \thead{Homogeneous,\\undirected}& \thead{Modulation\\recognition} & \thead{ Radio signal\\modulation\\ classification} & \thead{AvgNet} & \thead{Supervised,\\CE loss} & \thead{
 Classification\\ accuracy,\\F1 score, recall,\\ model size} & \thead{N/A} \\
  \cline{2-9}
   & \cite{Cammerer_gc_22} & \thead{Bipartite,\\
directed} & \thead{Channel\\decoding\\for LDPC} & BER min. & \thead{BGNN} & \thead{Supervised,\\ BCE loss} & \thead{Optimality} & \thead{Codeword bit} \\
   \hline
\multirow{13}{*}{Network}& \cite{Rusek_jsac_20} & \thead{Heterogeneous,\\directed}& \thead{
Routing\\optimization} & \thead{Delay, jitter,\\drop prediction} & \thead{RouteNet} & \thead{Supervised} & \thead{MRE} & Router \\  
 \cline{2-9}
  & \cite{zhang_tvt_25} & \thead{Heterogeneous,\\directed}& \thead{Routing\\optimization\\for LEO} & \thead{Delay min.} & \thead{GRLR} & \thead{DRL} & \thead{Optimality} & \thead{N/A} \\  
 \cline{2-9}
  & \cite{zhou_tnnl_24} & \thead{Homogeneous,\\directed} & \thead{Lightweight\\anomaly\\detection} & \thead{
  Attack\\classification} & \thead{RG-GLD} & \thead{Supervised,\\KD} & \thead{Classification\\
accuracy,\\F1-score, recall,\\false alarm rate,\\ detection rate,\\model size,\\inference time} & \thead{N/A} \\ 
 \cline{2-9}
  & \cite{song_nl_25} & \thead{Homogeneous,\\directed} & \thead{Routing\\optimization\\for FANETs} & \thead{Weighted \\path selection} & \thead{GNNPPOR} & \thead{PPO} & \thead{Throughput,\\energy\\consumption,\\delivery ratio,\\jitter} & \thead{N/A} \\
\hline
\end{tabular}
\label{SectionIII}
\end{table*}

In general, conventional wireless systems involve numerous graph-structured topologies in both the network edge and the network core. Most existing studies treat these topologies as homogeneous node-centric graphs and  employ GCN- or GAT-based models to handle related tasks.

The primary goal of GNNs is to enable high-performance and real-time processing, given that conventional algorithms based on iterative frameworks are computationally intensive and struggle to balance the trade-off between computational efficiency and optimization performance \cite{learing-to-optimize}. Moreover, driven by the escalating demand for high-data-rate traffic and the proliferation of IoT devices, wireless communication systems are evolving toward large-scale, distributed, and flat architectural paradigms. This trend further poses substantial challenges to network optimization, stemming from the exponential growth in problem complexity. Data-driven GNNs can be trained offline using historical data or prior knowledge, thereby facilitating efficient online communication and networking designs. Consequently, GNNs have emerged as a transformative tool to revolutionize wireless system designs, enabling the replacement of single or multiple conventional algorithms. 

Currently, the core focus of GNN-enabled wireless designs lies in infusing intelligence across diverse protocol layers (including the physical, media access control (MAC), and network layers) to enhance system performance \cite{8884151, 8382166}. Empowered by agentic GNNs, GNNs tailored for different layers can be coordinated to achieve a shared objective, thereby breaking the constraints of the protocol stack and realizing joint cross-layer gains. This section summarizes the advanced applications of GNNs as intelligence enablers in conventional wireless communication systems.

\subsection{GNN-Enabled Physical and MAC Layer}

\begin{figure}[t]
    \begin{center}
        {\includegraphics[ width=1\linewidth]{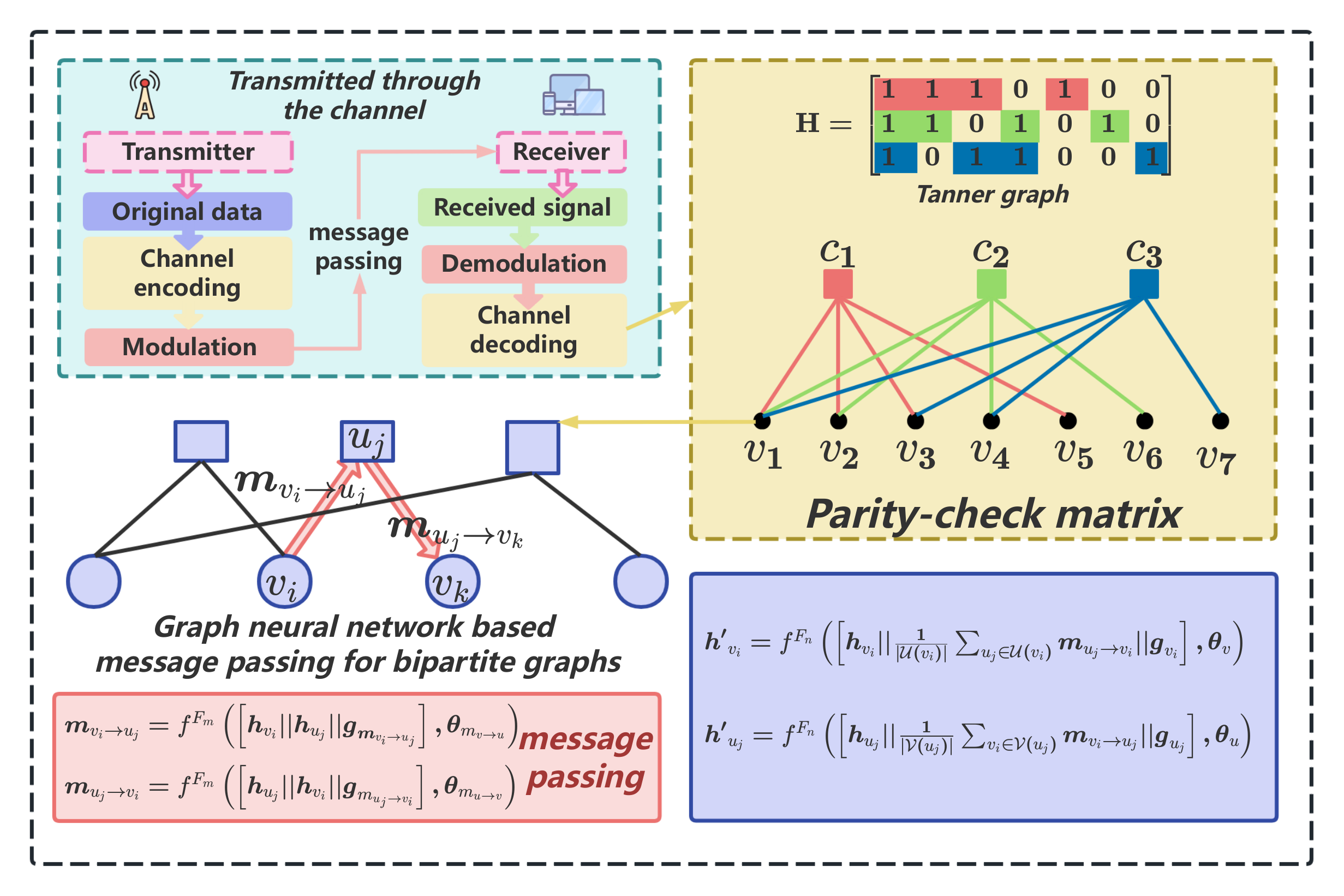}}
        \caption{Illustration of GNN-enabled channel decoding for LDPC codes. The Tanner graph, derived from the parity-check matrix, is represented as a bipartite graph consisting of check nodes and variable nodes. Message passing is performed to update the log-likelihood ratio assigned to each codeword bit \cite{Cammerer_gc_22}.}
        \label{mac}
    \end{center}
\end{figure}

\begin{figure}[t]
    \begin{center}
        {\includegraphics[ width=1\linewidth]{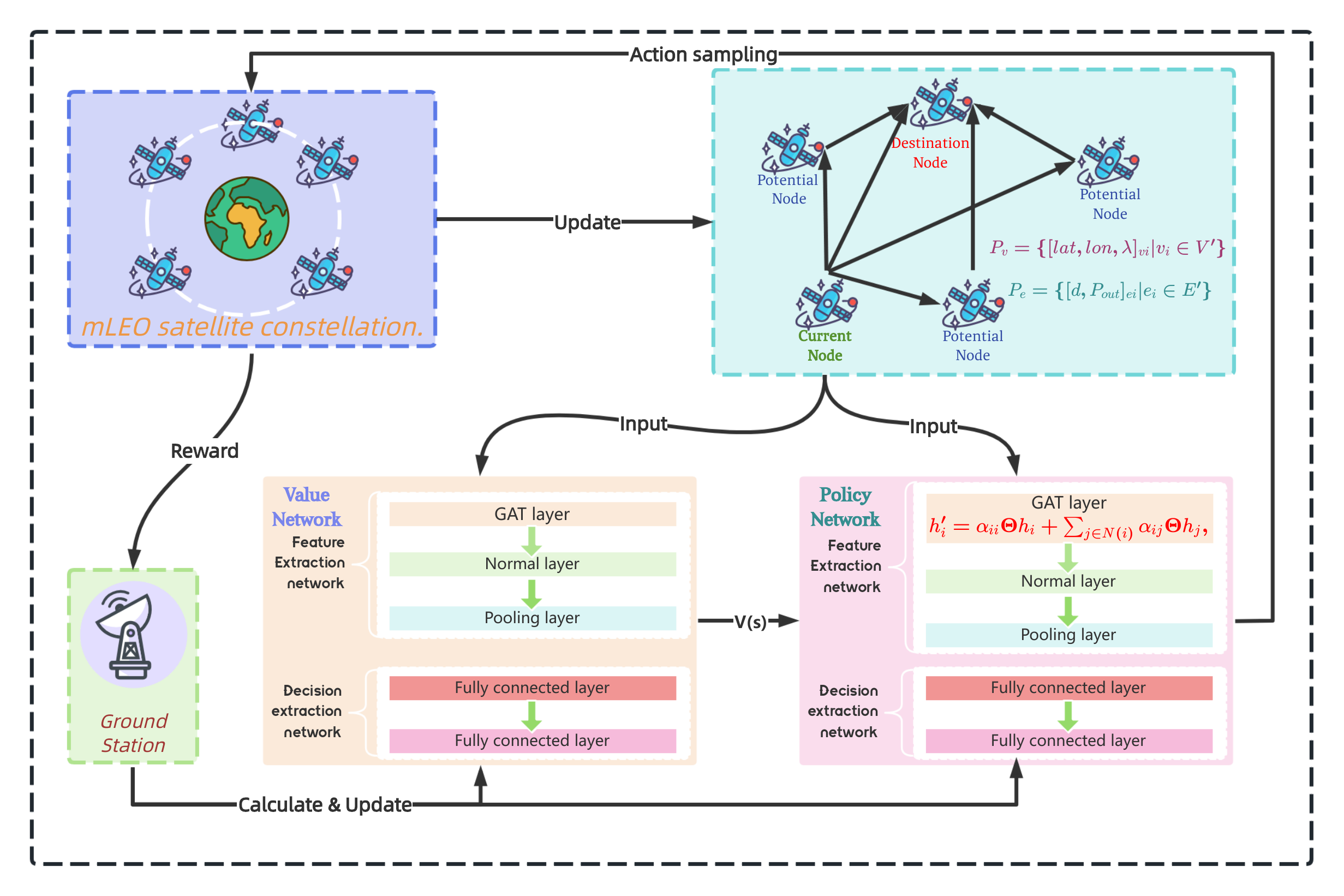}}
        \caption{Illustration of the GRLR routing scheme for mLEO satellite constellation. Structurally, it is built on the actor-critic RL framework to establish a decision network. Meanwhile, it leverages GNN to construct a feature extraction network, modeling information from satellites and inter-satellite links (ISLs) as a graph. GNN aggregates features from neighboring satellites and ISLs to update the satellite's features, enabling effective processing of heterogeneous network data\cite{zhang_tvt_25}.}
        \label{routing}
    \end{center}
\end{figure}

\begin{figure}[t]
    \begin{center}
        {\includegraphics[ width=1\linewidth]{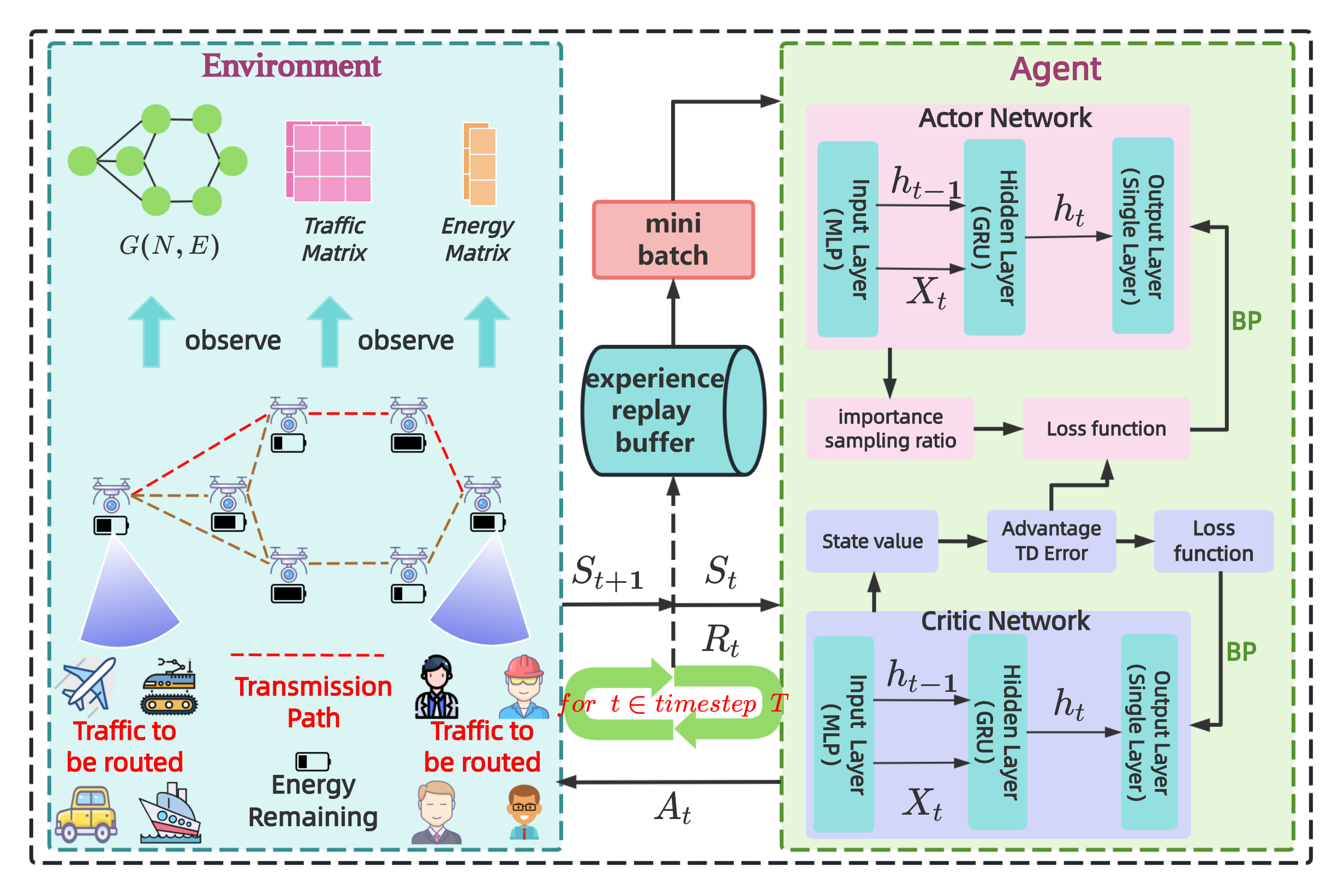}}
        \caption{The GNNPPOR architecture for FANETs routing. It uses the PPO framework with an actor-critic model. A graph model represents the FANET topology, traffic, and energy matrices. GRU-based GNN layers process node states, aggregate features, and update network status to generate routing paths. The GNN refines node features from environmental data, aiding routing decisions and state evaluation, with updates occurring via environmental interactions \cite{song_nl_25}.}
        \label{FANETs}
    \end{center}
\end{figure}

GNN-enabled designs for physical and MAC layers encompass resource allocation, beamforming design, channel estimation, channel decoding, and signal detection. 

\subsubsection{Resource allocation and beamforming design}
The throughput-optimal network node deployment problem is investigated in \cite{yang_tnnls_24}, where the authors propose a GNN-based approach incorporating graph convolutional layers to achieve a more precise throughput approximation and develop a GNN-based proximal policy optimization (PPO) \cite{schulman2017proximal}  algorithm for training data collection. Ablation studies demonstrate that GraphConv outperforms GINEConv \cite{hu2019strategies} and GCN convolution in capturing spatial dependencies; Gaussian error linear unit (GELU) \cite{hendrycks2016gaussian} outperforms rectified linear unit (ReLU) and LeakyReLU in modeling non-linear relationships; and global-add-pooling is superior to global-mean-pooling and global-max-pooling for feature aggregation. In \cite{li_twc_24_1}, a GNN-based beamforming design is proposed to maximize the sum rate under QoS and power budget constraints for multi-user multiple input single output (MU-MISO) networks, leveraging a GAT with complex-valued computations and residual connections. The model employs the penalty method and Lagrangian duality to handle QoS constraints, achieving near-optimal and real-time beamforming designs through unsupervised training, with scalability to the number of users and adaptability to varying power budgets, outperforming GNN-based baselines. Additionally, transfer learning is employed to enhance the model's scalability to unseen numbers of users. 

In addition to non-robust designs, the authors in \cite{liang_jsac_2024_0612} employ a bipartite GNN (BGNN) to address outage-constrained beamforming design under Gaussian channel state information (CSI) errors. They construct an empirical mapping from estimated CSI to a model-based robust beamforming expression. An MU-MISO system is modeled by a bipartite graph with antenna and user nodes, and fed into a BGNN where each layer comprises user message generation, antenna message generation, and feature decision. Regarding the satisfaction of the outage constraint, a data augmentation-based quantile estimation module is proposed to formulate an unsupervised loss function based on the pre-defined outage probability. Numerical results demonstrate that the approach achieves non-conservative robust performance, yielding higher data rates,  better power efficiency, and faster inference than some conventional optimization methods\cite{10071987},\cite{6891348}. In \cite{he_tvt_2025_0616}, the authors leverage GNNs to design beamforming vectors for optimizing rate outage-constrained energy efficiency (EE) in interference channels with statistical CSI. They model the interference channel as a directed fully-connected graph, where nodes and edges encode the communication and interference links, respectively. The GNN is built upon the equivariant GNN (EGNN) \cite{keriven2019universal}, a structure that preserves specific transformational equivariance (e.g., translation and rotation), and utilizes the attention mechanism to achieve robust beamforming design. Via unsupervised learning, the GNN achieves millisecond-level inference time compared to the second-level inference time of a block successive upper bound minimization (BSUM)-based algorithm, and incurs less than $5\%$ average optimality loss.

In \cite{Gu_twc_24}, the authors address a distributed beamforming and power control problem for massive ultra-reliable and low-latency communication (mURLLC) networks, aiming to minimize the logarithmic average utility of decoding error probability for the worst link. They propose two GNN models, termed graph for users (G4U) and probabilistic G4U (PG4U), both employing a one-layer GNN structure where graph embeddings are updated using the previous frame's states and current CSI. G4U performs GraphConv by broadcasting pilot signals to aggregate neighbor messages via amplitude/phase information, while PG4U determines policies using only previous-frame CSI, enabling multi-layer perceptron (MLP) execution during data transmission. Both frameworks use unsupervised training with a novel loss function based on the asymptotic expression of the Gaussian Q-function. Experimental results show significant advantages: G4U and PG4U outperform  GNN baselines\cite{Shen_jsac_21}, equal power allocation, and weighted minimum mean square error (WMMSE) \cite{shi2011iteratively} in the probability of QoS outage. Specifically, PG4U excels in short frame durations with high channel correlation, while G4U is suitable for moderate frames with low correlation.  In \cite{hu_TVT_24_0718}, the authors propose an unsupervised power allocation based on the attentive graph representation (UPAR) method in ad-hoc wireless networks. The approach integrates two core components: an edge aggregated graph attention regression network (EGRN) and a deep unfolded WMMSE (DU-WMMSE) module. EGRN leverages graph attention mechanisms and GRU units to aggregate node and edge features, enhancing the expressiveness and generalization of the learned graph representation. DU-WMMSE serves as a knowledge-driven solution readout module. The model effectively reduces the solution space of the considered problem, accelerating convergence and improving training efficiency. 

In \cite{Deng_jsac_25}, the authors investigate GNN-enabled hybrid analog and digital beamforming (HBF) for massive MIMO and mmWave systems, proposing a Sub-6G information aided multi-user hybrid beamforming (SA-MUHBF) framework to mitigate heavy pilot overhead in CSI acquisition. The framework consists of three stages: mmWave beamspace representation prediction, analog beam selection, and digital beamforming. Particularly, a 2D CNN predicts the mmWave beamspace representation from Sub-6G channel estimates, a multi-layer GNN architecture iteratively refines the quality of analog beam selection, and a linear minimum mean squared error (LMMSE) algorithm is used to design digital beamforming. For the GNN stage, the system is modeled as an interference graph, where each node represents a browser/server user link with beamspace features, and directed edges share attributes with their source nodes.  A novel graph convolutional layer integrates input preprocessing and 2D CNN-based feature convolution. SA-MUHBF is trained in two steps: supervised training for the prediction network, followed by unsupervised training for the GNN. Numerical results show SA-MUHBF outperforms the baselines \cite{maschietti2019coordinated} in sum rate, achieving $0.5$$\sim$$22.2\%$ gains across different numbers of users. The GNN’s ability to coordinate inter-user interference is highlighted, demonstrating robust performance under varying Sub-6G pilot powers, antenna configurations, and unseen scenarios.

In \cite{Yang_tvt_24_2}, the authors address a joint admission and power control problem for massive connection scenarios, aiming to maximize the number of links meeting QoS requirements via signal quality metrics (e.g., signal-to-noise ratio (SNR)) under blocking and accumulative interference models. For the blocking interference model, they utilize a GCN with a modified degree-adaptive loss function, where the penalty term is inversely correlated with node degree. The GCN performs message passing through iterative neighborhood aggregation, followed by post-processing to resolve conflicting nodes. For the accumulative interference model, a GAT is employed, using a $3$-head attention mechanism to aggregate neighbor features with adaptive weights and a sigmoid-activated fully-connected layer to output normalized power control values.  Simulations show that the GCN-based method achieves $83.8\%$ of the optimal solution within short runtime, outperforming fixed-penalty GCN \cite{schuetz2022combinatorial} and heuristic algorithms \cite{kianpisheh2021joint} in activated link count, especially for large node degrees. The GAT-based method improves activated links by $43.78\%$ over linear programming (LP) baselines  \cite{liu2012joint} and $10.37\%$ over unmodified GAT, with $10$$\sim$$100\times$ lower computational complexity than convex optimization methods. Both models demonstrate stability across varying interference thresholds and transmitter-receiver distances, with GAT also achieving energy savings.

\subsubsection{Channel estimation, channel decoding and signal detection}
In \cite{Liu_twc_25}, the authors propose a GNN-enabled joint channel estimation and data detection (JCD) method based on the bilinear generalized approximate message passing (BiGAMP) framework \cite{BiG-AMP} and signal conditional correlation, termed GNN-BiGAMP. This method integrates message passing into the data-detection loop by leveraging channel-symbol coupling. A training scheme combining pre-training and multi-task learning is also proposed, and the model achieves superior performance in symbol error rate (SER) and normalized mean square error (NMSE) compared to traditional and DL-based JCD methods \cite{ma2020joint, zhang2022message}. Further, they propose BiGNN-BiGAMP, which employs two GNNs to assist data detection and channel estimation. Trained via step-by-step supervised learning, the model outperforms existing models \cite{9018199} while requiring fewer pilots. Moreover, GNNs have demonstrated promise in channel decoding applications. 

In \cite{Cammerer_gc_22} and \cite{Gong_ISIT_24}, two GNN-based channel decoders are proposed for low-density parity-check (LDPC) codes. The former employs a BGNN as an end-to-end decoder, while the latter introduces a vanilla GNN with unrolled feedback as an intermediate perturbation adder between two consecutive belief propagation (BP) iterations. Both models are designed in a lightweight manner to ensure low decoding complexity. Numerical results evaluate the two models in terms of bit error rate (BER). The GNN-based decoders can achieve better performance than traditional BP decoders, and scale well to longer codeword lengths. 

In \cite{Xuan_tnse_22}, the authors tackle the task of radio signal modulation classification by developing an end-to-end GNN framework, termed AvgNet, which integrates the adaptive visibility graph (AVG) algorithm and a modified DiffPool model \cite{ying2018hierarchical}. As a variant of the visibility graph (VG) \cite{lacasa2008time}, AVG employs 1D convolution layers with ReLU activation to adaptively map time-series signals into graphs, constructing feature matrices where edge weights encode node correlations. DiffPool, a hierarchical graph pooling method, processes graph structures for classification while adhering to the message-passing paradigm. The framework is trained in a supervised manner using cross-entropy (CE) loss, with hyperparameter tuning applied to convolution kernel ranges to balance local and global feature extraction. Experimental results demonstrate that AvgNet outperforms RNNs (e.g., GRU, LSTM), CNNs (e.g., gramian angular field (GAF), Markov transition field (MTF) \cite{wang2015encoding}, 1D CNNs (CNN1D) \cite{o2018over}, and 2D CNNs (CNN2D) \cite{o2016convolutional}), as well as traditional VG methods (e.g., VG and limited penetration VG (LPVG)) in terms of accuracy, F1-score, and recall. Ablation studies validate  the critical role of adaptive local feature aggregation in AVG.

In \cite{Liu_tap_23}, the authors address the task of predicting the root-mean-squared (RMS) electric field strength distribution in highly variable indoor environments. The model comprises three main components: graph modeling, a GNN, and a predictor. In the graph modeling stage, a 3D geometry and transmit antenna are mapped to a graph with physical and unphysical edges. The GNN stage utilizes multiple hidden layers with message-passing operations: each hidden layer employs an MLP to fuse neighborhood information and edge attributes, directly outputting node embeddings. The final graph-level vector representation is obtained via global additive pooling. The predictor stage, fully-connected layers with residual links, serves as a decoder, combining the vector representation and observation point coordinates to predict the RMS electric field strength. Via supervised training with MSE loss, the GNN model demonstrates several advantages over traditional methods (e.g., in situ measurements, ray tracing algorithms) and other DL models (e.g., MLP and CNN). It exhibits exceptional generalization for fixed geometries, maintains robustness against irregular configurations in unknown geometries, and enables strong generalizability to arbitrary setups.

\subsection{GNN-Enabled Network Layer}

GNN-enabled designs for the network layer primarily aim to enhance routing protocols, particularly in intelligent traffic engineering (TE), including performance prediction, routing generation, and anomaly detection. Notably, the application of agentic AI for controlling and managing networks has attracted growing attention \cite{clark2003knowledge, 9651930}.

\subsubsection{Performance prediction}
The authors of \cite{Rusek_jsac_20} propose RouteNet, a GNN-based model for Knowledge-Defined Networking (KDN) \cite{mestres2017knowledge}, to accurately predict key performance indicators (KPIs). RouteNet is capable of learning and modeling graph-structured information and generalizing across arbitrary topologies and network settings. RouteNet can be regarded as a type of extension of a vanilla MPNN, and an RNN is adopted to aggregate link states along paths in order to capture the sequential dependence of links. RouteNet is shown to outperform the queuing-theoretic baseline developed in \cite{kelly2011reversibility} in terms of the mean relative error (MRE) of delay, jitter, and packet drops. Additionally, RouteNet can generalize effectively to topologies with varying nodes while remaining accurate in its estimates. 

A GNN-based model, termed RouteNet-Fermi (RouteNet-F), is proposed in \cite{Ferriol_TON_23}  for the purpose of constructing accurate data-driven network models \cite{chou2021survey} for predicting key network metrics (e.g., delay, jitter, and packet loss) in a network. Compared with queuing theory, RouteNet-F dispenses with the Markovian traffic assumption; compared with MLPs and RNNs, RouteNet-F is applicable to network scenarios unseen during the training stage. RouteNet-F is built upon a vanilla MPNN. RouteNet-F is evaluated in both simulation and real-world scenarios, and has achieved accuracy comparable to computationally intensive packet-level simulators while significantly reducing inference time.

\subsubsection{Routing generation}
Path-based GNN (PathGNN) is proposed in \cite{ye_jsac_25} to efficiently infer robust and resilient routing strategies in a distributed fashion, with the objective of enhancing the ability to handle unexpected traffic fluctuations and unpredictable link failures. By considering the task nature of TE (e.g., resource contention among traffic flows), PathGNN adopts a path-link bipartite graph to model paths and links as graph entities. The architecture incorporates the multi-head attention mechanisms and the skip connections. Trained via supervised learning guided by optimal routing solutions, PathGNN demonstrates superior generalization in unseen traffic scenarios and concurrent link failure conditions, compared to multi-agent RL-GNN \cite{MARL-GNN} (a GNN-based TE model under a multi-agent RL framework) and traditional protocols. Besides, PathGNN supports distributed implementation, facilitated by GNN function sharing among homogeneous graph entities. 

In \cite{song_nl_25}, the authors propose a GNN-based approach for PPO routing (GNNPPOR) to select routing paths in flying ad-hoc networks (FANETs). The primary objective of GNNPPOR is to enable effective network load distribution and QoS compliance. UAVs are represented by a directed graph, where nodes correspond to UAVs and edges carry FANET-specific information (including link load, traffic size, remaining energy, and bandwidth). GNNPPOR utilizes an actor network and a critic network to select and evaluate actions based on the observed  FANET state. These two networks are identical, employing  a GRU-based GNN  model where the GNN aggregates multi-dimensional network state information, followed by a GRU to update the network state. 

In addition to routing schemes for  terrestrial systems, some works have considered GNN-enabled routing for space-air-ground integrated networks. In \cite{zhang_tvt_25}, the end-to-end delay minimization routing problem in mega low Earth orbit (LEO) satellite constellations is formulated as a Markov decision process (MDP) and addressed by a GNN integrated with an RL algorithm, named GRLR. The mega LEO network is modeled by a weighted directed graph, where nodes and edges represent satellites and inter-satellite links, respectively. GAT is employed to extract the features from the graph representation and integrated into an actor-critic RL  framework to determine the next-hop action. GRLR is centrally trained by ground stations and distributively implemented  by satellites. 

\subsubsection{Anomaly detection}
GNNs have emerged as a powerful tool for network anomaly detection by leveraging their ability to model complex topological dependencies and extract structural-semantic features from network data. In \cite{zhao_tmc_24}, the authors address the fault scenario identification (FSI) task for communication networks, aiming to recognize fault types from massive alarms automatically. They propose a Knowledge Enhanced (KE)-GNN model that integrates the advantages of both the rules and GNNs. The system is modeled as a fault scenario graph, where nodes represent devices or alarms, and edges encode physical/logical connections. Knowledge is encoded using propositional logic and mapped into a knowledge space, with a teacher-student scheme designed to minimize the distance between knowledge embeddings and GNN predictions. Trained in a supervised manner on labeled real-world 5G fault datasets, KE-GNN achieves up to $99.15\%$ accuracy on FSD1-15, outperforming rule-based methods by $8.10\%$ and traditional GNNs by $5.41\%$-$9.83\%$. It demonstrates superior performance on small datasets and new carrier sites, highlighting strong generalization and robustness. Ablation studies validate the critical role of the teacher-student scheme in enhancing complex fault identification, while the hierarchical knowledge representation method proves effective for handling three levels of fault scenarios. 


In \cite{zhou_tnnl_24}, the authors address anomaly detection in IoT  networks, aiming to ensure secure data exchange in resource-constrained environments. The proposed RG-GLD integrates GNN and KD, using a GAT for structural feature extraction and an MLP for traffic feature analysis. It employs a graph network reconstruction strategy: data communications are modeled as nodes in a directed graph with edges defined by transmission rules, combined with a graph attention-based KD scheme that preserves lightweight local subgraphs while aligning global information. The training strategy consists of two stages: supervised training of the teacher model using graph and flow data, followed by knowledge transfer to the student model via combined losses. RG-GLD outperforms several baselines in classification accuracy and computational efficiency. It achieves lower false alarm rates and higher detection rates in anomaly detection, validated by receiver operating characteristic curves and confusion matrices. The effectiveness of GAT's multi-head attention and self-attention for global alignment is validated.

\subsection{Lessons Learned}
GNN-enabled designs in conventional wireless communication systems effectively balance the trade-off between computational efficiency and effectiveness through offline training. Most existing studies demonstrate that unsupervised and supervised learning achieve comparable performance, further alleviating the burden of collecting labeled datasets {\cite{gnn_mag}. Thus, leveraging GNNs to enable performance enhancement across diverse network layers and application scenarios should be a central focus in wireless intelligence research.

\section{GNN-Enabled Multifunctional  Services and Diverse Transmission}

\begin{table*}[t]
\centering
\footnotesize
\caption{GNN-Enabled Multifunctional  Services and Diverse Transmission.}
\begin{tabular}{|c||c|c|c|c|c|c|c|c| }
\hline
{\bf System} &  {\bf Ref.} & {\bf Graph} & {\bf Task} & {\bf Problem} & {\bf Model}   & {\bf Train}  &{\bf Metrics} &  {\bf Scalability}        \\ \hline\hline
\multirow{11}{*}{ISAC} & \cite{Zhao_twc_25} & \thead{Heterogeneous,\\ fully-connected,\\ 
 undirected} & \thead{Joint transmit \\
and sensing\\ beamforming design} &\thead{QoS-constrained \\ beampattern \\ MSE min.}  & GNN-IP   & \thead{Unsupervised,\\penalty terms,\\projection\\module} & \thead{Optimality,\\ scalability,\\feasibility rate,\\inference time} & UE  \\    
 \cline{2-9}
  & \cite{mao_jsac_25_0615} &\thead{Heterogeneous,\\ fully-connected,\\ 
 undirected} &\thead{Joint transmit \\
and sensing \\beamforming design} &\thead{QoS-constrained \\ CRB min.}& ISACNet&\thead{Unsupervised,\\penalty terms} & \thead{Optimality,\\feasibility rate,\\inference time} & N/A  \\ 
 \cline{2-9}
  & \cite{li_jsac_23_0615}& \thead{Heterogeneous,\\ 
 directed} & \thead{Multi-label\\classification\\in V2V} & \thead{Joint service mode\\selection and target\\ vehicle association} &  HGNN & Supervised & \thead{Classification\\ accuracy,\\feasibility rate,\\inference time} & N/A \\  
 \hline
SWIPT & \cite{han_tmc_25} & \thead{Homogeneous,\\ fully-connected,\\ 
 undirected } & \thead{Joint beamforming\\and PS/TS\\ratio design} & \thead{QoS-constrained \\ sum-rate  max.}  & GAT   & \thead{Unsupervised,\\ penalty terms,\\transfer learning} & \thead{Optimality,\\ scalability,\\feasibility rate,\\Inference time} & UE  \\     
\hline
\thead{MISO}  & \cite{songz_tvt_24} & \thead{Heterogeneous,\\ 
 undirected} & \thead{Beamforming design\\for PLS}  & \thead{Secure sum \\rate max.\\ \& EE max.} & HGNN & Unsupervised & \thead{Optimality,\\ scalability,\\feasibility rate,\\Inference time} &\thead{Eve,\\ UE}\\
 \hline
\multirow{5}{*}{UAV}& \cite{krishnan_tvt_25}& \thead{Homogeneous,\\fully-connected,\\
undirected} & \thead{Covert\\communication} & \thead{Detectability\\min.} & GKAE  & \thead{Self-supervised,\\multi-task} & \thead{Prediction\\error,\\detection\\probability}  & N/A\\
 \cline{2-9}
 & \cite{tang_tmc_25_0618} & \thead{Homogeneous,\\ fully-connected,\\ 
 undirected} & \thead{Joint beamforming \\ design and UAV \\placement for PLS} & \thead{Secure sum \\ rate max.}& \thead{DGRL} & \thead{SAC,\\transfer learning} & \thead{Optimality,\\inference time} & N/A \\
  \cline{2-9}
\hline
\thead{Edge\\networks}  & \cite{Yu_jsac_23} & \thead{Heterogeneous,\\undirected} & \thead{Service self-healing} & \thead{Performance\\ prediction,\\abnormality \\detection} & \thead{DGAT,\\DGMA} & \thead{Supervised} & \thead{Prediction\\accuracy,\\load balance\\degree, resource\\utilization} & \thead{Edge\\nodes}\\ 
\hline
\end{tabular}
\label{SectionIV}
\end{table*}

\begin{figure}[t]
    \begin{center}
        {\includegraphics[ width=1\linewidth]{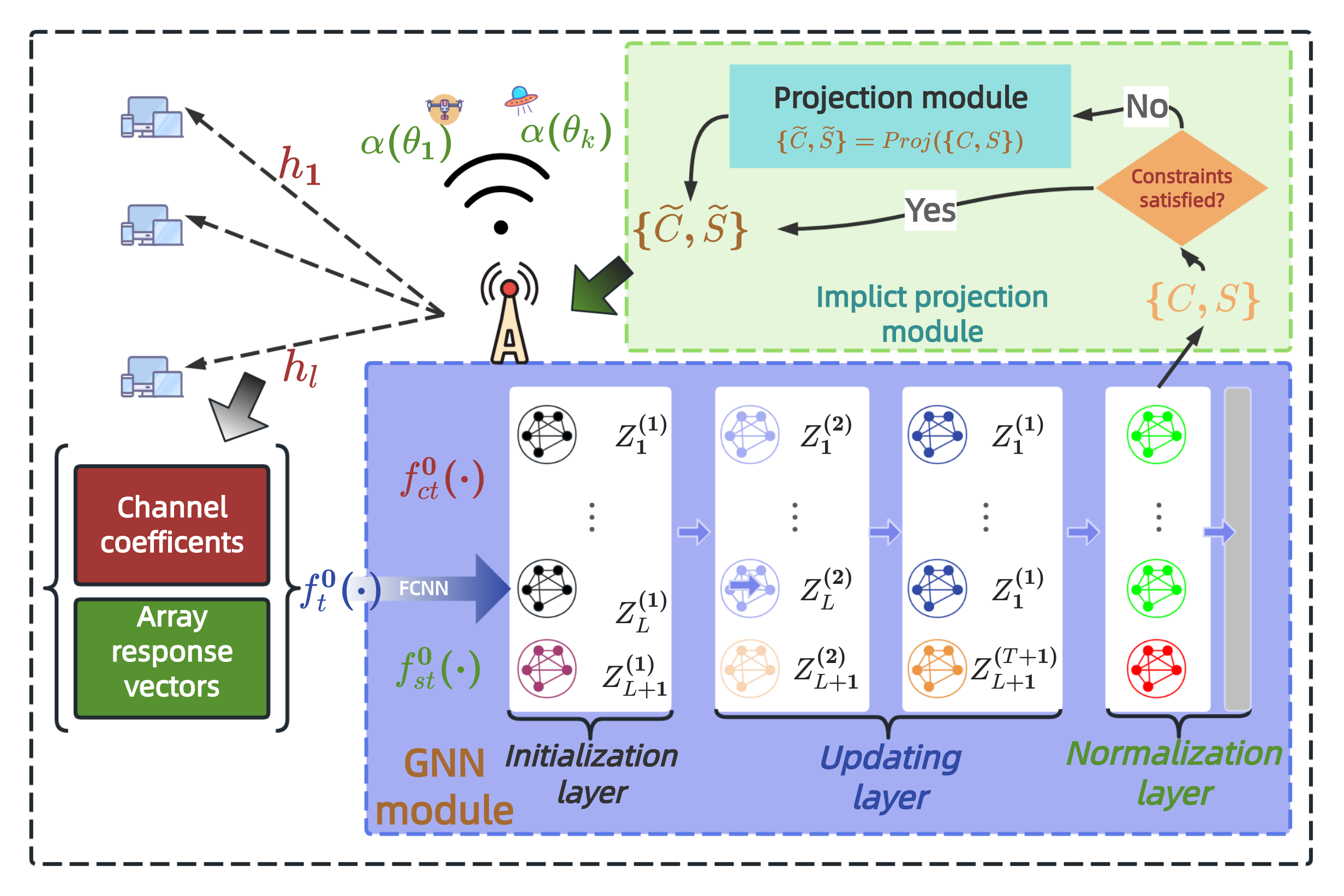}}
        \caption{The workflow of the GNN-IP algorithm comprises two main modules. The GNN module maps the CSI of communication users and array responses of sensing targets to coarse beamforming matrices. Then, the projection module refines these obtained coarse beamforming matrices to meet the communication requirements \cite{Zhao_twc_25}.}
        \label{isac}
    \end{center}
\end{figure}

\begin{figure}[t]
    \begin{center}
        {\includegraphics[ width=1\linewidth]{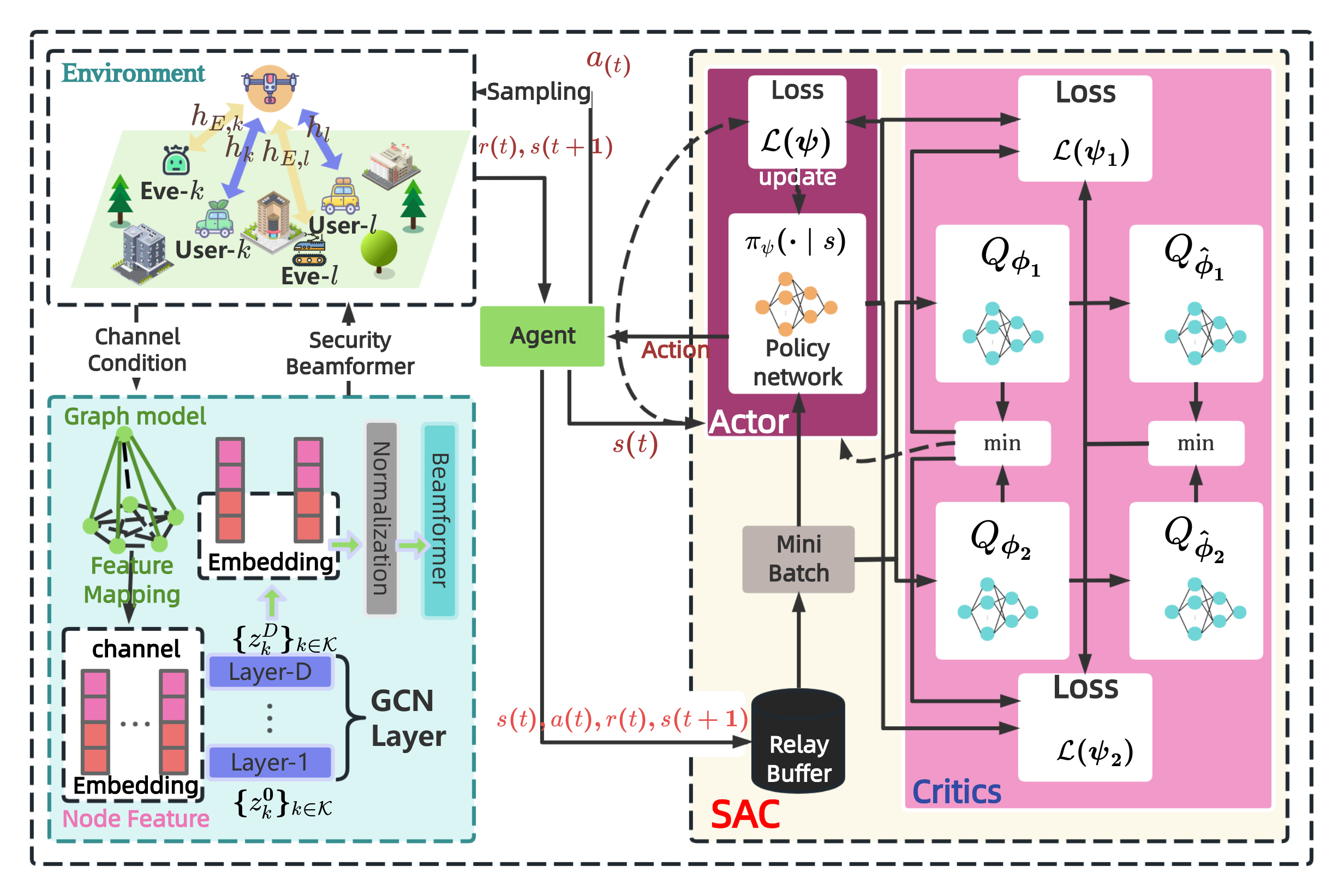}}
        \caption{Illustration of the DGRL architecture for secure beamforming in UAV communications. The system employs the SAC framework with dual critics. A graph model is constructed to represent communication nodes, and GCN layers are utilized for node embedding and feature processing to generate beamformers. The GNN model is updated through environmental interactions to optimize UAV placement by iteratively refining the mapping between node features and beamforming strategies\cite{tang_tmc_25_0618}.}
        \label{uav}
    \end{center}
\end{figure}

Unlike traditional wireless systems focused on information transmission, multifunctional services and diverse transmission scenarios aim to utilize radio frequency (RF) signals to simultaneously achieve multiple objectives. Since GNNs need to represent complex interactions and handle heterogeneity, HGNNs are enabled to find their applications.


The provision of multifunctional services has emerged as a pivotal evolutionary trend for mobile communication systems, enabled by the abundant spectrum available in high RF bands. For example, information-bearing RF signals can simultaneously serve as radar sensing waveforms in ISAC systems and energy carriers in simultaneous wireless information and power transfer (SWIPT) systems\cite{6489506}. A theoretical challenge in multifunctional system designs lies in the unification of performance metrics for different services across diverse spatial and temporal dimensions. Building upon prior theoretical analyses, researchers have extensively investigated performance metrics beyond traditional information transfer. For instance, various radar sensing performance metrics have been proposed for ISAC, including the Cramér–Rao bound (CRB) \cite{liu2021cramer}, sensing signal-to-interference-plus-noise ratio (SINR), and signal-clutter-noise ratio (SCNR), facilitating  joint resource allocation for information transmission and radar sensing tasks. With clearly defined performance metrics, GNNs have been leveraged to enable real-time and scalable resource allocation. A key technical challenge lies in addressing the heterogeneity across network entities, particularly in multifunctional scenarios. This challenge stems from the fact that sharing a single GNN-based feature extractor across different feature types can significantly undermine the model's representational capability.

Similarly, RF signals can establish diverse link types, such as uplink/downlink links, information/leakage links, and desired/interference links, which exhibit inherent heterogeneity. Designing GNNs to distinguish these link types enables finer-grained feature extraction, effectively integrating prior knowledge about link characteristics. 

In light of the multifunctional and diverse RF signals, agentic GNNs can play a crucial role in autonomously assigning tasks to appropriate GNNs or coordinating multiple GNNs to perform service handover. This section summarizes the advanced applications of GNNs in realizing the multifunctional and diverse RF signal propagation.

\subsection{GNN-Enabled Multifunctional Services}

GNNs have been leveraged for ISAC services in cellular and vehicular networks and have empowered SWIPT.

\subsubsection{ISAC in cellular networks}
 In \cite{Zhao_twc_25}, the authors propose a learning-to-beamform framework for ISAC. They consider an ISAC scenario involving multiple communication users and sensing targets, aiming to develop an efficient and scalable algorithm that optimizes the radar transmit beampattern subject to the communication performance. They propose a GNN with implicit projection (GNN-IP) approach: the GNN maps CSI to transmit beamforming matrices, and an implicit projection module ensures solution feasibility. The system is modeled as a graph, and user nodes and the  target node are initialized via distinct features, and updated through different aggregation and combining operations.  GNN-IP is trained using an unsupervised loss function based on a Lagrangian transformation. Compared with a semidefinite relaxation (SDR)-based algorithm \cite{luo2010semidefinite}, GNN-IP achieves a similar sensing MSE while significantly reducing computational complexity and maintaining stable runtime. Ablation experiments show GNN-IP maintains $100\%$ feasibility under varying SINR and user counts, whereas other DL models exhibit reduced feasibility under stricter constraints or increased user scales.
 
 In \cite{mao_jsac_25_0615}, the authors present an attention-based framework for bistatic ISAC systems, comprising a transmit BS, a receive BS, downlink/uplink users,  and targets. They first derive closed-form expressions for the CRB for target parameter estimation and formulate the CRB minimization problem subject to downlink/uplink communication requirements, transmit power budgets, positive semi-definite requirements for sensing beamformers, and unit-modulus constraints for receive beamformers. To solve this problem, they develop two approaches: a successive convex approximation (SCA)-based method for stationary-point solutions and a learning-based ISACNet for real-time resource allocation. The ISACNet architecture employs a pre-processing module to transform CSI into high-dimensional node features, an attention-based encoder with multi-head self/cross-attention to capture complicated intra/inter-node interference patterns, and a decoder generating beamforming solutions while satisfying  power constraints, positive semi-definiteness, and unit-modulus requirements. Notably, communication rate constraints are incorporated as penalty terms in an unsupervised training loss function. Simulations demonstrate that ISACNet achieves $>90\%$ performance of the SCA method, enabling millisecond-level responses and $1000\times$ computational efficiency gains. The aforementioned GNN-enabled ISAC designs enable efficient collection of environmental information. Further, the EGI can utilize task-specific AI models to process sensing data (typically fused with other modalities or source information), thereby facilitating intelligent applications.
 
\subsubsection{ISAC in vehicular networks}
In addition to traditional mobile networks, some studies have focused on  GNN-enabled ISAC for vehicular networks, as intelligent vehicles emerge equipped with directional antennas for communication and short-range radar. In \cite{Lee_tvt_22}, the authors investigate a vehicular system with vehicles traveling along a dual-carriageway. Their goal is to develop a joint optimization scheme for the application layer encompassing radar detection, transmission data, served vehicles, and resource allocation. They model vehicle interactions via MDP and propose a GNN-based multi-agent DRL approach. Vehicles and communication links are modeled as nodes and edges, respectively. The nodes (also agents in DRL) exchange and update node features via GCN. The output features are then fed into an action decoder and a value decoder of multi-agent PPO \cite{9601214}, yielding the agents' radar or communication actions. Experimental results demonstrate that the approach significantly outperforms non-learning baselines in communication efficiency and radar detection accuracy, while exhibiting robustness against varying vehicle counts and environmental conditions. 

In \cite{li_jsac_23_0615}, the authors study the vehicular ISAC systems,  aiming to maximize the sum rate of the communication  vehicles, while satisfying the sensing service requirements of the target vehicles. The optimization problem involves joint design of service modes and vehicle association.  The system is modeled as a heterogeneous graph comprising  three  node types: service provider vehicles, communication vehicles, and sensing target vehicles. Besides, the links connecting nodes are defined based on both source and destination node types. Leveraging the graph representation, each node performs link-aware message passing for feature updating, followed by feature concatenation to derive the final solution. On datasets \cite{8446749}, simulation results demonstrate that the proposed HGNN-based scheme achieves $93\%$ of the sum rate of  the exhaustive search algorithm, outperforming a homogeneous GNN-based algorithm \cite{9322537} and a conventional optimization method. Further, the authors extend their work to the joint communication and sensing optimization problem for terahertz (THz) vehicular networks under blockage and mobility conditions in \cite{li_twc_24_2}. They formulate an optimization problem to maximize the total number of successfully served vehicles by jointly optimizing the service mode and vehicle association. A dynamic GNN model is proposed to select appropriate graph information aggregation functions according to the vehicle network topology. The model utilizes a supervised learning strategy with binary CE (BCE) loss and mini-batch stochastic gradient descent (SGD) for training. Numerical results demonstrate that the dynamic GNN-based method enhances the number of successfully served vehicles by up to $17\%$ and $28\%$ compared to a fixed-model GNN algorithm and a conventional non-GNN optimization approach, respectively.

\subsubsection{SWIPT} SWIPT is also a kind of multifunctional service. In \cite{han_tmc_25}, a unified GNN-based model was proposed for SWIPT networks, termed SWIPTNet. Here, the ``unified" indicates that SWIPTNet is able to handle the transmission design for both time-switching (TS) and power-splitting (PS) receivers. Transfer learning was adopted to further enhance the learning performance. The two types of services, i.e., information and energy transfer, are unified by a single-type output approach, where the PS or TS ratio is represented by a function of beamforming vectors. SWIPTNet is built upon GAT, and the Laplacian transformer and layer connection are adopted to augment the model's expressive capability. SWIPTNet is numerically shown to achieve near-optimal performance across diverse scenarios, and outperform GCN and vanilla GAT.

\subsection{GNN-Enabled Diverse Transmission}

Existing studies have endeavored to differentiate among various link types within GNN-based frameworks, leveraging GNNs to model their heterogeneous characteristics and interrelationships.

\subsubsection{Information security and privacy}
In \cite{songz_tvt_24}, the authors adopt an HGNN to design physical-layer secure  beamforming in MISO systems. The heterogeneity between legitimate users and eavesdroppers (Eves) is considered. Specifically, the HGNN categorizes the interactions between users and Eves into three distinct types, i.e., inter-user interference, information leakage from users to Eves, and inter-Eve interaction, and thus, partitions the graph into three subgraphs with each dedicated to one link type. The multi-head attention and the semantic attention \cite{you2016image} are used respectively to capture the type-specific features within subgraphs and cross-subgraph feature fusion. The model is trained unsupervisedly to address secure sum-rate and EE maximization problems in a unified manner. The performance gains in optimality and scalability from heterogeneous processing are evident when compared to CVX-based algorithms and CNN-based \cite{huang2019fast} baselines. 

In \cite{tang_tmc_25_0618}, the authors leverage deep GRL (DGRL) for secure beamforming in UAV-enabled multi-user communications. They formulate a secure sum rate maximization problem subject to transmit power constraints and UAV flight area limitations. The system is modeled as a graph where each node represents a legitimate user, with features constructed by concatenating the user's CSI, the eavesdropper's CSI, and an auxiliary embedding. The GNN architecture incorporates multiple GCN layers, followed by normalization operations on updated embeddings to generate feasible beamforming vectors. UAV deployment is derived via a soft actor-critic (SAC)  RL algorithm \cite{haarnoja2018soft}, where the GNN-based secure beamforming module guides iterative updates of the deployment strategy. 


In \cite{Lee_twc_24}, the authors address the privacy-preserving decentralized inference problem in wireless networks, where information exchanges via wireless channels may cause privacy leakage. Their goal is to enhance inference privacy using GNNs while maintaining performance. They propose a GNN with a layer-wise structure: each layer comprises local message, aggregation, and update functions implemented via MLPs, following the general message-passing framework. The training strategy employs unsupervised learning, using the negative sum rate as the loss function. The over-the-air (OTA) computation technique is analytically and experimentally shown to enhance communication/computation efficiency and privacy, further reinforcing the model's effectiveness in decentralized inference scenarios.  An SNR-privacy trade-off analysis shows the proposed approach achieves an optimal balance between inference accuracy and privacy.


\subsubsection{Covert communication}
In \cite{krishnan_tvt_25}, the authors address the task of enabling predictive covert communication in terrestrial ad-hoc networks under multi-UAV surveillance, aiming to predict UAV trajectories for transmit power optimization and detectability minimization. They propose a Graph Koopman Autoencoder (GKAE), where the GNN component models the multi-UAV network as a time-varying graph with node features and adjacency matrices defined by spatial proximity. The GNN employs standard message-passing operations to generate compact graph embeddings capturing spatial interactions, which are then fed into a Koopman autoencoder to linearize non-linear dynamics in a latent space for long-term trajectory prediction. The model is trained via a two-step self-supervised strategy, i.e.,  the GNN encoder-decoder is optimized to reconstruct graph structures   and the Koopman autoencoder \cite{lusch2018deep} is trained to predict future states using graph embeddings. Experimental results show that GKAE achieves significantly lower detection probability and higher trajectory prediction accuracy than traditional DL baselines \cite{kant2022long, cho2014learning}. 

In \cite{hao_iot_25_0618}, the authors propose  a general steganographic framework for NNs based on GCNs, with the objective of enabling covert communication by embedding secret data during network training. The framework transforms hidden layers in NNs into graph structures, where nodes encode network entities (e.g., linear layer features, convolutional kernels) and edges represent predefined relationships, thus facilitating data embedding via GCN-based message passing. The training process is guided by a composite loss function combining mean squared error for embedding accuracy and the original task loss. GCN parameters can be either randomly initialized or explicitly specified, and graph connectivity is pre-established between the sender and receiver, thereby eliminating the need for parameter transmission. Experimental evaluations across image classification, segmentation, generation, and language processing tasks demonstrate that the framework achieves superior security and robustness compared to baseline methods \cite{wang2021data, yang2023general}. Pretrained GCNs further optimize parameter distributions, reducing Kullback-Leibler (KL) divergence by approximately $40\%$ and highlighting the efficacy of dynamic graph interaction modeling and parameter efficiency in steganographic applications.

\subsubsection{Network interaction construction} In \cite{Yu_jsac_23}, the authors utilize Digital Twin (DT) \cite{9170905} to enhance GNN-enabled wireless optimization and handle network heterogeneity. They investigate service self-healing in 6G edge networks, addressing challenges in abnormal node/link detection and service recovery under network heterogeneity and dynamic loads. A DT-driven mechanism leveraging a Dynamic GAT (DGAT) is proposed, which constructs a network interaction graph with three node types (links, queues, flows) and four relationship types (link-flow, queue-flow, flow-queue, queue-link). Dynamic attention mechanisms enable message passing to model complex component interactions. DGAT is trained in a supervised manner using flow-level delay predictions from packet-level simulator OMNet++ as labels, with MSE as the loss function. Experiments show DGAT outperforms GCN, GraphSAGE, and Transformer in key metrics. For service redeployment,they propose a GNN-based Deep Graph Matching Algorithm (DGMA) that employs bi-directional graph embedding and the Sinkhorn-Hungarian algorithm for node matching, reducing service delay and improving overall load balance compared to several baselines\cite{7907163,dijkman2009graph}.  Ablation studies validate the DT-driven proactive prediction framework: combining node/link anomaly prediction enhances load balance, and DGAT’s dynamic attention yields higher prediction accuracy than static GCN.

\subsection{Lessons Learned}
Both  multifunctional services and diverse transmissions introduce heterogeneity within networks. Existing works have identified two primary approaches to tackling  heterogeneity: one unifies diverse feature types into a homogeneous representation, enabling the use of conventional homogeneous GNNs for network optimization; the other employs heterogeneous learning to model distinct features, which is better suited for HGNNs. Moreover, HGNNs inherently offer scalability to accommodate an expanding variety of network elements. However, HGNNs may give rise to more complex neural architecture designs and lead to prolonged inference time.

\section{{GNN-Enabled Flexible Antennas and Cell-Free Architecture}}

\begin{table*}[t]
\centering
\footnotesize
\caption{GNN-Enabled Flexible Antennas and Cell-Free.}
\begin{tabular}{|c||c|c|c|c|c|c|c|c| }
\hline
{\bf System} &  {\bf Ref.} & {\bf Graph} & {\bf Task}& {\bf Problem} & {\bf Model}   & {\bf Train}  &{\bf Metrics} &  {\bf Scalability}        \\ \hline\hline
\multirow{16}{*}{RIS/IRS} & \cite{Singh_tgcn_24}  & \thead{Homogeneous,\\fully-connected,\\
undirected} & \thead{Cascaded channel\\estimation} & NMSE min. &  G-TIRC   & Supervised & \thead{Optimality,\\scalability,\\inference time}  & \thead{IRS\\element} \\
 \cline{2-9}
 \cline{2-9}
& \cite{dai_wcl_24} & \thead{Homogeneous,\\fully-connected,\\
undirected} & \thead{Channel \\extrapolation} & \thead{Element selection,\\NMSE min.} &  GCN  & \thead{Supervised,\\multi-\\objective} & Optimality& N/A \\  
\cline{2-9}
& \cite{Le_twc_25} & \thead{Bipartite,\\
undirected} & \thead{Joint active \\and passive \\beamforming design} & \thead{Sum rate \\max.} &  BHGNN  & \thead{Unsupervised,\\ Lagrangian\\ duality} & Optimality& \thead{RIS\\element,\\ UE} \\  
 \cline{2-9}
& \cite{Lyu_TCE_24}  & \thead{Heterogeneous,\\
fully-connected,\\ undirected} & \thead{Joint
 BS \\beamforming and \\RIS phase\\ configuration} & \thead{Sum rate \\max. w/o\\ channel estimation} &  HGNN  & \thead{Unsupervised} & \thead{Optimality,\\ scalability,\\training time,\\ inference time}& \thead{UE} \\  
  \cline{2-9}
& \cite{wang_TMC_25} & \thead{Heterogeneous,\\
fully-connected,\\ undirected} & \thead{Joint  beamforming \\and trajectory design\\for UAV} & \thead{Average sum rate \\max. and total\\  time slot min.} &  \thead{SD3-GNN-\\RIS}  & \thead{SD3} & \thead{Optimality, \\trajectory}& \thead{N/A} \\  
 \hline
FAS &  \cite{he_tvt_2025} & \thead{Homogeneous,\\fully-connected,\\
directed}  & \thead{Joint beamforming \\and antenna \\placement design}  &\thead{Sum rate max.\\ \& EE max.} & \thead{Two-stage \\
GAT}  & Unsupervised &\thead{Optimality,\\scalability,\\inference time}& UE \\  
 \hline
PA & \cite{guo_wcl_2025_0604} & \thead{Bipartite,\\ 
 undirected} &  \thead{Joint pinching \\and transmit \\beamforming design} &  \thead{Sum rate\\max.} &GPASS & Unsupervised & \thead{Optimality,\\scalability} & UE \\  
\hline
\multirow{14}{*}{Cell-Free}& \cite{Liu_WCNC_25} & \thead{Bipartite,\\ 
 undirected} & AP selection & \thead{Uplink
sum \\SE max.} & R-GCN & Unsupervised & \thead{Optimality,\\fairness,\\feasibility rate,\\inference time}& N/A \\
 \cline{2-9}
  & \cite{Ranasinghe_GC_21} & \thead{Bipartite,\\ 
 undirected} & \thead{AP association \\
transmit design} & \thead{Lower-bound \\sum rate max.} & \thead{MHSB-\\GNN} & Unsupervised & \thead{Optimality,\\FLOPs,\\scalability,\\inference time}& UE \\
  \cline{2-9}
  & \cite{tung_tvt_25} & \thead{Bipartite,\\ 
 undirected} & \thead{Distributed\\power allocation} & \thead{Sum ergodic \\ rate max.} & \thead{Distributed \\ GNN} & Unsupervised & \thead{Optimality,\\scalability,\\inference time}& AP, UE \\
   \cline{2-9}
   & \cite{Zhou_iot_25} & \thead{Homogeneous,\\fully-connected,\\
undirected} & \thead{ADC resolution\\allocation} & \thead{Uplink \\sum rate max.} & \thead{DMAGNN-\\AC} & \thead{Actor-critic } & \thead{Optimality}& N/A \\
\hline
\end{tabular}
\label{SectionV}
\end{table*}

\begin{figure}[t]
    \begin{center}
        {\includegraphics[ width=1\linewidth]{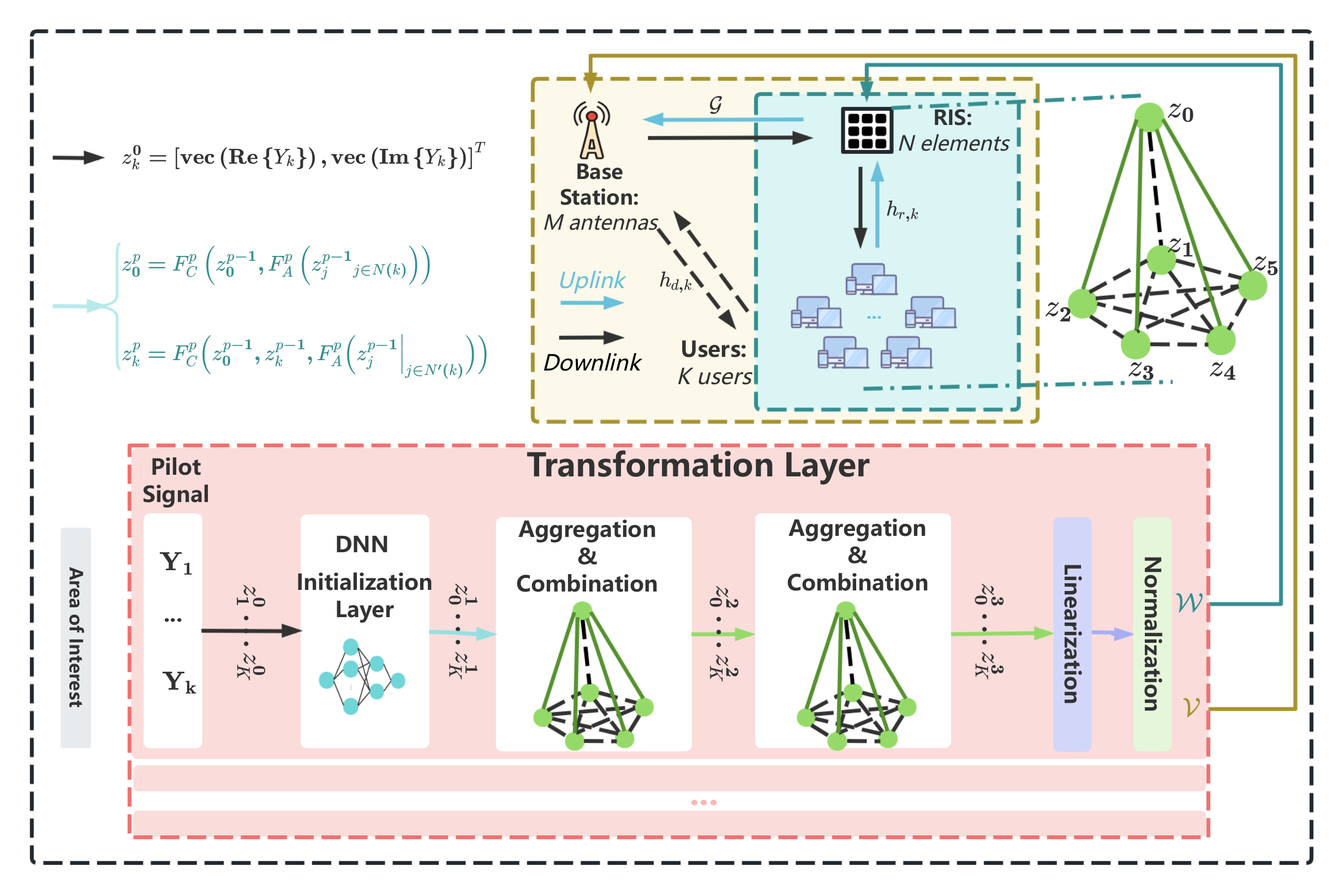}}
        \caption{Illustration of the HGNN framework for jointly optimizing BS beamforming and RIS reflection coefficients. The RIS and users are modeled as distinct node types. Heterogeneous message passing mechanisms are applied to each node type. The outputs of the HGNN architecture guide the transmit design of both the antenna array and the RIS\cite{Lyu_TCE_24}.}
    \end{center}
\end{figure}

\begin{figure}[t]
    \begin{center}
        {\includegraphics[ width=1\linewidth]{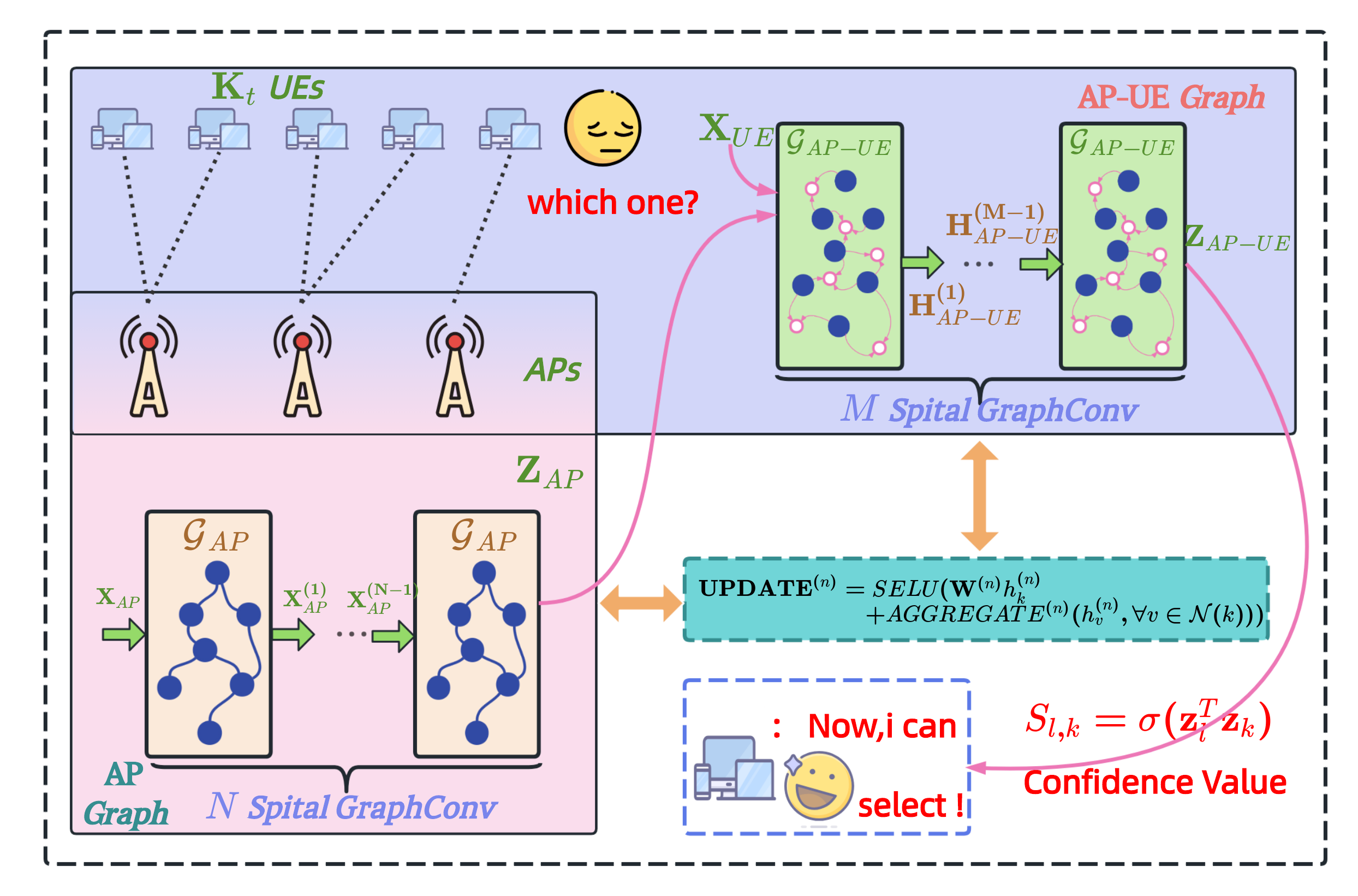}}
        \caption{Architecture of the confidence value-based MHSB-GNN AP selection method, which utilizes an AP graph to learn the spatial layout and structural relationships among APs, and an AP-UE graph for link prediction between APs and UEs \cite{Ranasinghe_GC_21}.}
        \label{ris}
    \end{center}
\end{figure}


As mentioned, the utilization of high-frequency spectrum (e.g., millimeter-wave and sub-THz bands) enables abundant available bandwidth and supports multifunctional wireless services\cite{9411894}. However, high-frequency signals are susceptible to severe propagation loss and adverse effects of environmental obstacles. Recently, the flexible-antenna technology and CFmMIMO  have emerged as promising solutions for future communication systems to enhance the propagation of high-frequency RF signals by customizing the user-centric wireless environment. For example, RIS/IRS adjusts the phase shifts of incident signals to establish new cascaded transmission links that go around obstacles; FAS or movable antennas enable small-scale antenna redeployment to alleviate the impacts of deep fading; pinching antennas (PAs)  facilitate  large-scale antenna deployment in close proximity to users to reduce path loss \cite{ding2024flexibleantennasystemspinchingantennaperspective}; CFmMIMO coordinates APs to achieve more uniform coverage. 

Although flexible antennas and CFmMIMO introduce new degrees of freedom, they present significant challenges to signal processing arising from deep coupling across multiple resource dimensions. Most existing studies on signal processing for flexible antennas and CFmMIMO have employed alternating optimization (AO) techniques, which often fail to achieve global optimality. Consequently, the hardware gains might not be fully exploited due to the lack of efficient solvers. Instead of building upon explicit mathematical formulations, DL enables  learning solutions  from data, and the well-trained DL models can serve as solvers for complex problems. As a result, many recent studies have sought to apply GNNs to signal processing tasks for optimizing the performance of flexible antennas and CFmMIMO. These models are more meticulously tailored than GNNs designed for conventional systems, aiming to address both the heterogeneity across network elements and the deep coupling of controllable variables.

In future communication systems, multiple technologies will be integrated into a single system, which necessitates the development of scenario-aware agentic GNNs to match GNN-enabled techniques with specific requirements. Moreover, novel techniques may lack sufficient data and performance labels, yet agentic GNNs can autonomously collect new data through environmental interaction and enable continuous performance improvement. This section summarizes recent GNN-enabled designs for flexible antennas and CFmMIMO.

\subsection{GNN-Enabled Flexible Antennas}

The GNN-enabled flexible-antenna designs primarily focus on two key technical frontiers: channel estimation and joint transmit design. 

\subsubsection{Channel estimation for RIS}
Channel estimation remains a challenging issue in novel flexible-antenna systems, particularly in RISs with multiple passive elements. In \cite{Singh_tgcn_24}, the authors address  the channel estimation problem for IRS-assisted systems  by leveraging the spatial proximity of reflecting elements. They propose a graph Transformer-based IRS channel estimation (G-TIRC) model. The approach first uses the least squares (LS) method to estimate channels for pre-divided IRS groups. G-TIRC, combining GNNs and Transformer architectures, then employs GNNs to extract inter-group channel correlations and Transformer layers to predict channels for unknown groups. Experimental results show that G-TIRC outperforms conventional methods\cite{9103231}, achieving significant improvements in estimation accuracy and reductions in training overhead. 

The authors in \cite{Ye_WCL_24} propose a GNN-based cascaded channel estimation method for RIS-assisted uplink MU-MISO systems. The graph representing the system comprises multiple user nodes initialized with received signals and location information, and an RIS node initialized by taking mean operation on user node features. The GNN employs MPNN layers to update user node features and a linear layer to yield the estimated channel matrix. A supervised normalized MSE (NMSE) loss function is used for training. Numerical results demonstrate that the GNN maintains robust performance and outperforms existing DL approaches\cite{9090876} under low pilot overhead, while generalizing effectively to varying uplink transmit powers. 

Besides, the authors in \cite{dai_wcl_24} propose a GCN-based channel extrapolation scheme for passive RIS systems without direct links. To reduce estimation cost, only selected elements are activated during the channel estimation stage, and CSI of inactive elements is extrapolated by the GCN. A fully-connected homogeneous graph is considered with each node representing an RIS element and featured by CSI.  The system is modeled as a fully-connected homogeneous graph, where each node represents an RIS element with CSI as features. The scheme comprises three key components: mapping full channels to a sparse graph, GCN-based node selection, and GCN-based channel extrapolation. Simulation results show that the proposed GCN effectively extracts spatial features at low sampling rates, enabling efficient pattern selection and channel extrapolation. 

\subsubsection{Joint beamforming for RIS}
Traditional AO methods often struggle with joint optimization of deep coupling variables, which can be addressed effectively by DL. In \cite{Le_twc_25}, the authors study a joint active and passive beamforming design for distributed simultaneous transmitting and reflecting (STAR) RIS-assisted multi-user MIMO systems. A sum-rate maximization problem subject to power budget is formulated, and addressed by an HGNN-based model, termed BHGNN. The system is represented by a heterogeneous graph, where the STAR-RIS elements and users form two node sets, and undirected edges connect RIS element nodes and user nodes. Nodes are initialized with beamforming vectors or phase shifts and amplitudes, while edges encode inter-node CSI features. Node features are updated via heterogeneous message-passing iterations, followed by a scaling operation to yield a feasible transmit design. A supervised Lagrangian loss function is adopted to train HGNN to achieve sum-rate performance comparable to an AO-based baseline, demonstrating strong generalizability across system configurations, including varying user numbers, STAR-RIS quantities, and element coefficients. 

In \cite{Chen_WCL_24}, the authors consider a downlink MISO RIS-aided multi-user system with rate-splitting multiple access (RSMA). The system is modeled as a graph comprising  RIS, private user, and common user nodes. RIS nodes are unconnected to each other but linked to private users via weighted edges; private user nodes are fully-connected by unweighted edges; the common user node connects to private user nodes through weighted directional edges. The GNN, based on MPNN, updates the node and edge features to generate RIS phase shifts from RIS nodes, beamforming vectors from user nodes, and common rate allocations from weighted edges. Via unsupervised training, the proposed GNN outperforms  DNN and SCA\cite{9145189} in EE and inference speed, achieving feasibility rates exceeding $99.5\%$ via the penalty method. 

In \cite{liu_TWC_25_002}, the authors propose an HGNN-based framework for joint beamforming design and RIS association in multi-RIS multi-user mmWave systems. A weighted sum-rate maximization problem is formulated for multi-RIS assisted systems. The graph representing the system includes three node types: BS, RISs, and users, which are connected by direct edges encoding CSI. The HGNN updates node features via MPNN layers tailored to node types. The beamforming vectors and reflecting element coefficients are generated from BS node and RIS nodes, respectively, while an RIS association scheme is yielded by a Softmax layer from user nodes. Numerical results demonstrate that the HGNN outperforms conventional DNNs by an order of magnitude, and an RIS association scheme improves system performance by approximately $30\%$.

In \cite{wang_TMC_25}, the authors integrate UAVs and RISs to enhance system flexibility and improve transmission efficiency in rich cluster environments. A hybrid algorithm, named soft deep deterministic policy gradient (SD3)-GNN-RIS, is proposed to maximize system rate while minimizing UAV energy consumption and flight duration. This algorithm combines the SD3\cite{9684973} for UAV trajectory optimization with a GNN for beamforming design. The graph representation and GNN updating mechanisms for a given time slot mirror those in \cite{Jiang_jsac_21}. Numerical results validate the algorithm's effectiveness in maximizing the achievable overall rate, with additional performance gains observed when multiple RIS-mounted UAVs collectively optimize beamforming and trajectories in MU-MISO systems. 

In \cite{wang_TWC_23}, the authors leverage  RIS to enhance the communication capability of OTA FL. They develop a GNN-based model to map CSI to transceivers and the desired RIS design minimizing time-average error, which remains independent of the number of edge devices. The proposed GNN algorithm comprises an initialization layer, multiple graphical mapping layers, and a parameter generation layer.  Particularly, three types of nodes, representing edge devices, edge server and RIS, execute aggregation and combination operations using features from all nodes. Extensive simulations demonstrate that  the GNN algorithm outperforms the AO-based algorithm\cite{1658226} in optimizing transceiver and RIS, offering lower complexity, higher training efficiency, and superior scalability. 

In addition to the one-stage  optimization paradigm, the two-stage approach has emerged as a critical research direction. In \cite{yang_Tvt_24}, the authors propose a precoding algorithm based on an unsupervised learning GAT to address the sum-rate maximization problem for RIS-assisted systems with blocked direct links. The first GAT predicts the RIS phase shift matrix using BS-RIS and RIS-user channel matrices to derive the effective channel matrix, which is then fed into the second GAT for generating the precoding matrix. The system is modeled as a fully-connected graph with user nodes in each GAT layer. Numerical results validate the GAT's learning capability and multi-head attention mechanism for sum-rate enhancement, demonstrating that the GAT outperforms methods such as  AO, vanilla GNN, CNN, and MLP.

\subsubsection{Joint channel estimation and transmission for RIS}
Some existing studies intend to integrate channel estimation and transmit design into an end-to-end learning paradigm. In \cite{Jiang_jsac_21}, the authors propose a DL approach for joint optimization of  beamformers and reflecting element coefficients, tailored to system-level objectives. This is achieved by using a GNN to parameterize the mapping from received pilot signals (incorporating user location information) to optimized system configurations. The graph includes user nodes and a single RIS node. The GNN comprises an initialization layer, multiple MPNN layers, and an output layer, aiming to transform user node features into feasible beamformers  and reflecting element coefficients. The results demonstrate that the GNN trained via unsupervised learning efficiently solves utility maximization problems with significantly fewer pilot signals than conventional methods\cite{8982186}. 

The authors of \cite{Yeh_wcl_24} address the limitations of two-step learning (i.e, channel estimation and transmit design) and end-to-end learning, and propose an Enhanced-GNN (E-GNN) approach for beamforming design in RIS-assisted mmWave systems with angular CSI. E-GNN comprises an enhanced feature extraction layer and  modified GAT layers. The graph representation and feature updating framework in \cite{Yeh_wcl_24} are similar to those in \cite{Jiang_jsac_21}, but the input is the estimated angular information from a learned approximate message passing (LAMP) network and enhanced by a 3D convolution layer \cite{lamp}, and the updating iteration is based on GAT. Numerical results validate the effectiveness of integrating the 3D convolution layer in the angular domain and the attention mechanism for suppressing inter-user interference. E-GNN shows stable performance and approaches the upper bound\cite{9087848} under different conditions. 

Furthermore, the authors in \cite{Lyu_TCE_24} address transmit design in large-scale RIS-assisted systems, while eliminating the need for channel estimation. They propose a GNN-based approach with region-specific training models. The graph for the system comprises user nodes and one RIS node, and all nodes are fully-connected. The received input pilots processed by a DNN are initialized as node features. Nodes undergo aggregation-combination operations followed by linearization and normalization layers to generate feasible beamforming vectors and reflecting coefficients. The authors further optimize regional layouts to prepare training datasets for area-specific GNNs. Experiments validate the approach, demonstrating accurate system performance with reduced complexity in large-scale communication scenarios.

\subsubsection{FAS and PA}

While most existing research on GNN-enabled flexible antenna systems focuses on RIS, a growing trend highlights the application of GNNs in FAS, movable antennas, and PAs. In \cite{he_tvt_2025}, the authors propose a GNN-based model to optimize the FAS system with the goal of EE maximization. The model comprises two stages, where the first is for antenna placement while the second is for beamforming design. The structures of the two stages are based on complex GAT, and their outputs are jointly fed into an unsupervised loss function corresponding to various system utilities. The effectiveness of the two stages is validated. 

In \cite{xie_wcl_2025_0604}, the authors propose a bipartite graph representation for a single-waveguide PA system, where the user nodes and antenna nodes are linked by undirected edges. The graph is fed into a bipartite GAT to solve an EE maximization problem via jointly optimizing antenna placement and power allocation. The updated node features are processed by two tailored readout modules to ensure feasible solutions. Numerical results validate the superiority of unsupervised bipartite GAT-enabled PA systems in terms of optimality, scalability, and computational efficiency, outperforming traditional CVXopt-based methods and systems.  In \cite{guo_wcl_2025_0604}, the authors propose a two-stage GNN-based model, termed GPASS,  for a multi-waveguide multi-user PA system, aiming to maximize the sum rate. The model is built for bipartite graphs, where two distinct sub-GNNs are employed to first learn the pinching beamforming and then the transmit beamforming. Numerical results show that the proposed architecture can achieve a higher spectral efficiency than the heuristic baseline method\cite{10595507} with low inference complexity.

\subsection{GNN-Enabled Cell-Free Massive MIMO}

The GNN-enabled CFmMIMO designs mainly focus on AP association, transmit design and distributed scheme. 

\subsubsection{AP association}
In \cite{Liu_WCNC_25}, an AP selection problem is formulated for CFmMIMO systems with the objective of maximizing the uplink sum SE. A heterogeneous GNN based approach via unsupervised learning is proposed to address the problem. The system is represented by a bipartite graph consisting of non-feature AP and user nodes, and undirected edges representing communication links connect nodes of distinct types with feature of large-scale fading coefficients. The heterogeneous GNN leverages the R-GCN framework \cite{R-GCN}, where messages from two directions (AP-to-user and user-to-AP) are aggregated separately and then combined. The updated node features for APs and users are concatenated and fed into an MLP to generate AP selection results. Numerical results demonstrate that the heterogeneous GNN achieves at least an 8\% average uplink sum spectral efficiency (SE) improvement over existing methods\cite{8097026, doan2018uplink, greening1990parallel}, with superior  fairness and scalability. The authors in \cite{Ranasinghe_GC_21} also focus on GNN-enabled AP selection for CFmMIMO systems. The system is represented by two graphs, i.e.,  a homogeneous graph with AP nodes and a heterogeneous graph with AP and user nodes. GraphSAGE is employed to update the node features which are then utilized to derive the confidence matrix of AP-user links. Numerical results demonstrate that the GNN-based AP selection outperforms proximity based AP selection algorithms, scales efficiently with the number of users, and reduces the requirement for reference signal receiving power measurements. 

The authors in \cite{Jiang_tvt_25} model CFmMIMO systems as a spatio-temporal graph, and employs  a spatio-temporal GNN (STGNN) to maximize a proposed coverage cost metric, named benefit-cost ratio (BCR), by optimizing AP locations. The  graph at a time slot comprises nodes representing APs, users, and coverage units, and edges connecting nodes. Due to the mobility, node and edge features vary across time and space.  The STGNN processes  the spatio-temporal graphs up to time slot $T$ to predict  AP deployment locations for time slot $(T+1)$, which leverages GAT for spatial learning and GRU for temporal learning, respectively.  Numerical results demonstrate that the proposed algorithm achieves $47.8\%$ and $7.8\%$ BCR improvements over random deployment and on-demand AP deployment baselines.

\subsubsection{Transmit design}
Some works intend to combine the AP association and transmit design to pursue a joint performance gain. In \cite{yan_twc_24}, the authors investigate the cross-layer optimization problem of AP selection and beamforming for CFmMIMO systems based on local CSI. The problem is first addressed by two CVXopt-based algorithms leveraging fractional programming and AO. Then, a lightweight multi-head single-body GNN (MHSB-GNN) algorithm is proposed. The system is modeled as a graph with beam and neighboring group  nodes, along with non-interference, intra-group, and inter-group interference edges. MHSB-GNN comprises an initial module, multiple MPNN-based node updating modules and a normalization module. Through unsupervised learning, MHSB-GNN outperforms the FP and the AO-based algorithms with significantly faster inference speed.  

In \cite{Mishra_twc_24}, the authors address  a max-min rate maximization problem in CFmMIMO systems with full or partial AP-user connectivity. The objective  is to map large-scale fading coefficients and AP-user associations to power control schemes. They utilize a directed heterogeneous graph to represent the system. Nodes denote the AP-user links with the initial features of large-scale fading coefficients, and edges capture inter-node relationships with two types and four attributes, defined by the source and destination node status. The proposed GNN is based on GAT, and trained with a supervised loss where labels are derived  by an second-order cone programming (SOCP)\cite{alizadeh2003second}. Numerical results show that the GNN achieves a median spectral efficiency within 4\% of the optimal SOCP baseline while reducing FLOPS by 10 to 100 times. The approach also demonstrates strong generalizability across deployment sizes, radio propagation environments, and per-AP serving densities.

In \cite{Li_twc_24_3}, the authors address a power allocation problem for maximizing SE  in a distributed multicarrier-division duplex (MDD) CFmMIMO system. They model the system as a heterogeneous graph with  user and AP nodes, each of which is associated  with two metapaths regarding communication  and interference links. Node features incorporate channel gain, power budget, and interference levels (including cross-link and self-interference), while edge attributes encode Euclidean distances. The proposed HGNN, termed CF-HGNN, integrates adaptive node embedding layers to handle varying node feature dimensions, metapath-based message passing to aggregate information from communication  and interference  paths with prioritized weights, and a metapath attention mechanism to learn path-wise information importance. Trained in an unsupervised manner with a customized loss function enforcing QoS  and power budget feasibility, CF-HGNN achieves $99\%$  SE of the quadratic transform and SCA (QT-SCA) algorithm with only $10^{-4}$ of the computational time and outperforms greedy allocation  methods\cite{jang2003transmit} and traditional DL models (including deep NN (DNN)\cite{labana2021unsupervised} and CNN\cite{bashar2020deep}). Ablation validates the impact of metapath attention and adaptive embedding layers, showing that structured message passing and node-type specific processing are critical for handling heterogeneous network dynamics. 
\subsubsection{Distributed scheme}
Distributed design in CFmMIMO is also critical, it enables seamless coverage of large areas with numerous distributed APs. In \cite{tung_tvt_25}, the authors propose a distributed GNN design for power allocation in CFmMIMO systems, aiming to maximize the sum ergodic rate. The system is represented by multiple graphs, each corresponding to one AP. Nodes and edges represent AP-user links and inter-link interference, respectively. The GNN, based on MPNN, is trained centrally on a CPU and shared to all APs for distributed inference. Numerical results show that the distributed GNN design achieves performance comparable to centralized one and outperforms traditional optimization-based approaches.

To model mobility and dynamic wireless environments, GNNs are increasingly integrated with DRL, enabling adaptive decision-making in time-varying communication systems. In \cite{liu_twc_25_0627}, the authors investigate the joint mobility and downlink power control in CFmMIMO systems equipped with mobile APs. A sum-rate maximization problem is formulated under constraints of inter-AP distance, movement area, and power budget. To capture the mobility of APs, the problem is transformed into an MDP, and handled by an MARL-based approach incorporating GNN to improve scalability and collaboration,. The CFmMIMO system is modeled as a graph comprising AP and user nodes, and then, processed through several MPNN layers to enable agents (e.g., APs and users) to update their self-observations.  

The authors in \cite{Zhou_iot_25} address uplink sum-rate maximization for CFmMIMO systems with non-ideal hardware components, such as nonlinear power amplifiers  and low-resolution analog-to-digital converters. They propose a GNN-assisted actor-critic algorithm named  DMAGNN-AC and its enhanced variant for power allocation, where each graph node represents a user with CSI as node features. A bilinear attention mechanism is employed  to calculate the inter-node association strengths to reflect  complex interference patterns. To mitigate computational complexity, attention coefficients are quantized to binary values, enabling efficient multi-scale feature aggregation.  Numerical results show that  DMAGNN-AC doubles the uplink sum rate of the full-power allocation baseline.

\subsection{Lessons Learned}
Flexible antenna and CFmMIMO communication systems diverge from conventional setups in both systemic architecture and algorithmic design. The former poses challenges in channel estimation due to complex propagation environments, while the latter suffers from theoretical intractability in guaranteeing optimal solutions. Based on existing works, we have two key observations for GNN-enabled flexible antennas and CFmMIMO. First, employing GNNs facilitates  novel communication paradigms, e.g., merging channel estimation and transmit design into an end-to-end learning paradigm to eliminate the need for explicit  channel estimation. Second, GNNs with unsupervised learning can achieve both better performance and faster inference than  CVXopt-based algorithms, e.g., AO. Notably, as system complexity escalates, developing efficient GNN-based approaches is challenging, particularly in the graph representation and neural architecture.

\section{SurveyLLM for GNN Applications}

There is a growing number of GNN applications in wireless systems, and agentic GNNs can autonomously utilize GNNs to tackle complex wireless tasks and reduce human intervention. However, the proliferation of GNNs may make it challenging for  operators and readers to quickly grasp their key points. Thus, alongside agentic GNNs, there is also a requirement to  facilitate easy access to GNN-related references.

While the primary objective of a survey is to provide a fundamental overview of emerging wireless technologies, such surveys often cover extensive content and references—making it time-consuming to locate answers to specific questions. Typically, readers must manually navigate through sections, search through references, or organize content to derive insights from these surveys. This is because traditional approaches to reading surveys rely on unidirectional information flow limiting their adaptability to individual research needs and impairing responsiveness to rapidly evolving fields.

\subsection{Advantages of SurveyLLM}
In light of LLMs as a powerful intelligent tool for knowledge reasoning, we propose a novel framework, termed SurveyLLM \cite{arslan2024survey}. This framework leverages LLMs as dynamic interfaces for accessing surveys, enabling enhanced comprehension and interactive engagement. By ingesting and synthesizing a corpus of surveys and references to form a knowledge base, SurveyLLM can transform the user experience from passive reading to active conversational exploration. For example, readers can pose natural-language queries (e.g., "\emph{Compare methods X and Y for task Z}" or "\emph{What are recent advances in concept A since paper B?}"), receiving synthesized and context-aware responses as well as relevant references. Specifically, the advantages of SurveyLLM over traditional reading approaches are summarized as follows.
\begin{itemize}

    \item \emph{Query-Centric Retrieval.} On-demand retrieval of relevant information from the knowledge base, tailored to readers' specific questions.  

    \item \emph{Reference  Searching.} Return and insert relevant references (with accessible links) into the answers as supporting documentation.
    
    \item \emph{Flexible Exploration.} Flexible yet comprehensive exploration of diverse concepts and their interrelationships.  
    
    \item \emph{Multi-Source  Synthesis.} Real-time summarization, comparison, and clarification of concepts across multiple sources.  

    \item \emph{Idea Provocation.} Guided dialogue that imitates readers' thinking process to provoke novel ideas.

\end{itemize}

SurveyLLM transcends the limitations of static text, facilitating readers' access to the knowledge base constructed from the survey. The semantic understanding and generative capabilities of LLMs enable the framework to function not only as a search engine to improve reading efficiency, but also as an analytical tool to provide actionable insights.

\subsection{Technical Implementation of SurveyLLM}


The goal of SurveyLLM is to build a Q\&A application based on LLMs and the local knowledge base. The technical implementation workflow is illustrated in Figure \ref{SurveyLLM}, which is  built on  the RAG \cite{lewis2020retrieval} framework and comprises a knowledge base, a classification model, and LLMs. The detailed processes of each component are given as follows.

The knowledge base is built upon the original documents (e.g., this article and its references) and a corresponding vector database generated by an embedding model (e.g., \cite{chen2024bge}). The documents are first partitioned into semantic chunks to facilitate fine-grained retrieval. Subsequently, each chunk is processed by the embedding model, which converts textual information into high-dimensional vector representations that preserve semantics. Finally, the vectors are indexed in the vector database to enable efficient approximate nearest-neighbor (ANN) searches\cite{johnson2019billion}. This module enables quick location of the most relevant chunks for the readers' queries (processed by the embedding model) in terms of semantic similarity. 

The classification model (e.g., \cite{BERT}, \cite{ml_knn}) operates on both original documents from the knowledge base and queries from readers. The former categorizes the fragments separated from the documents into thematic clusters and assigns multiple labels to each fragment following the organization of this article. Particularly, the labels may include the publication year, subsection title (of this article), graph representation, GNN architecture, etc. The latter labels queries to match them with their most relevant fragments. The process enhances retrieval precision by filtering results using specific fragments with the same category labels.

LLMs deployed locally are built upon specific open-source LLM architectures, such as DeepSeek \cite{deepseekai2024deepseekv3technicalreport}, LlamaGPT \cite{touvron2023llama}, etc., aiming to deliver synthetic answers anchored in references. These answers are generated using prompts designed via templates (such as few-shot \cite{reynolds2021prompt} and chain-of-thought (CoT)  \cite{wei2022chain}) under the RAG framework. Particularly, these prompts integrate key information from users' queries,  document fragments identified by the classification model, and chunks from vector databases retrieved via ANN search \cite{white2023prompt}. Such an approach enhances the domain-specific responses, as the prompts incorporate both context-aware semantic understanding and task-oriented structural information. Furthermore, LLMs can be fine-tuned via parameter-efficient techniques (e.g., LoRA \cite{hu2022lora}) to strengthen  their alignment with domain-specific knowledge.

\begin{figure} [t]
	\subfloat[\label{1a}]{
		\includegraphics[width=0.48\textwidth]{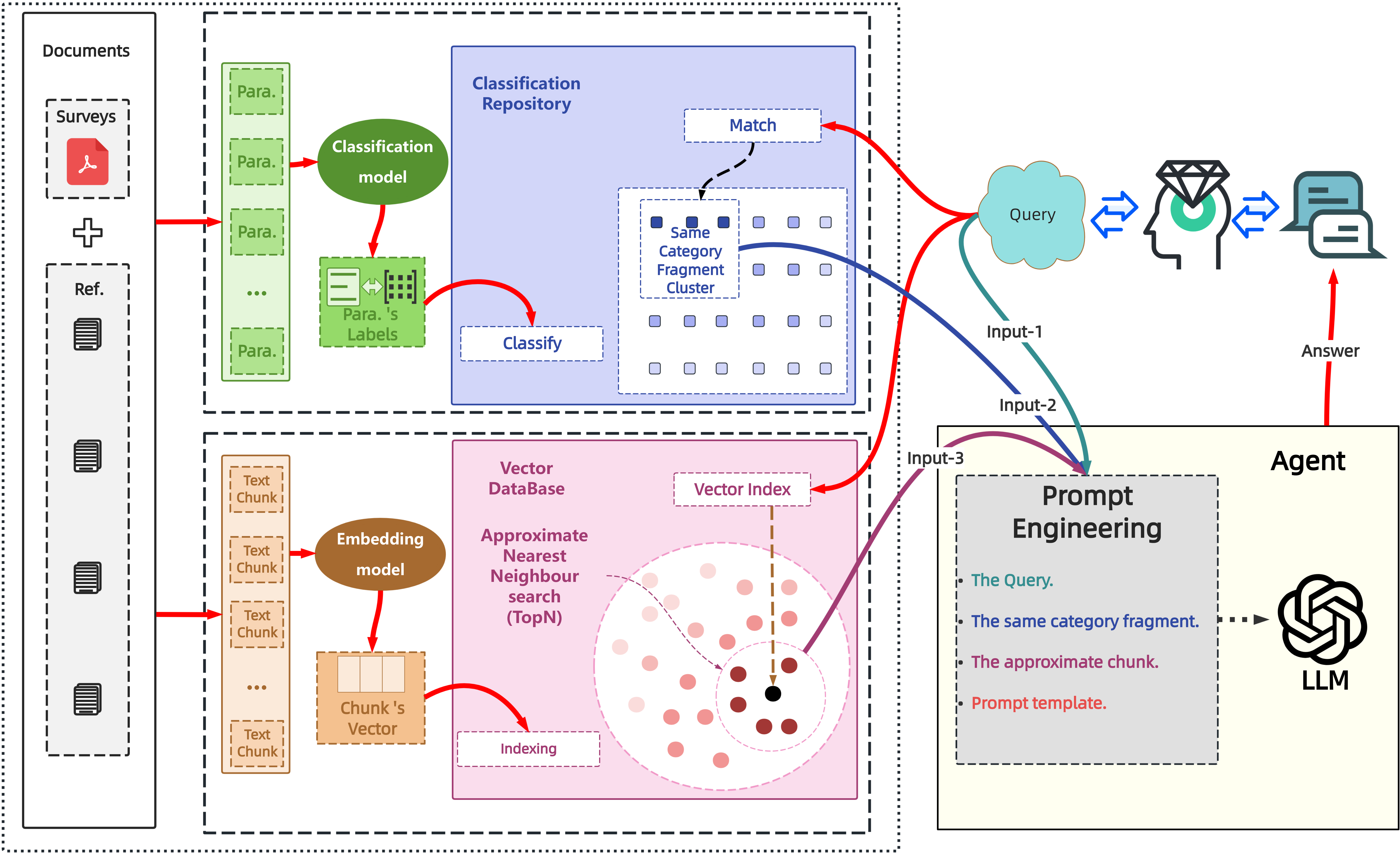}}\\
	\subfloat[\label{1b}]{
		\includegraphics[ width=0.48\textwidth]{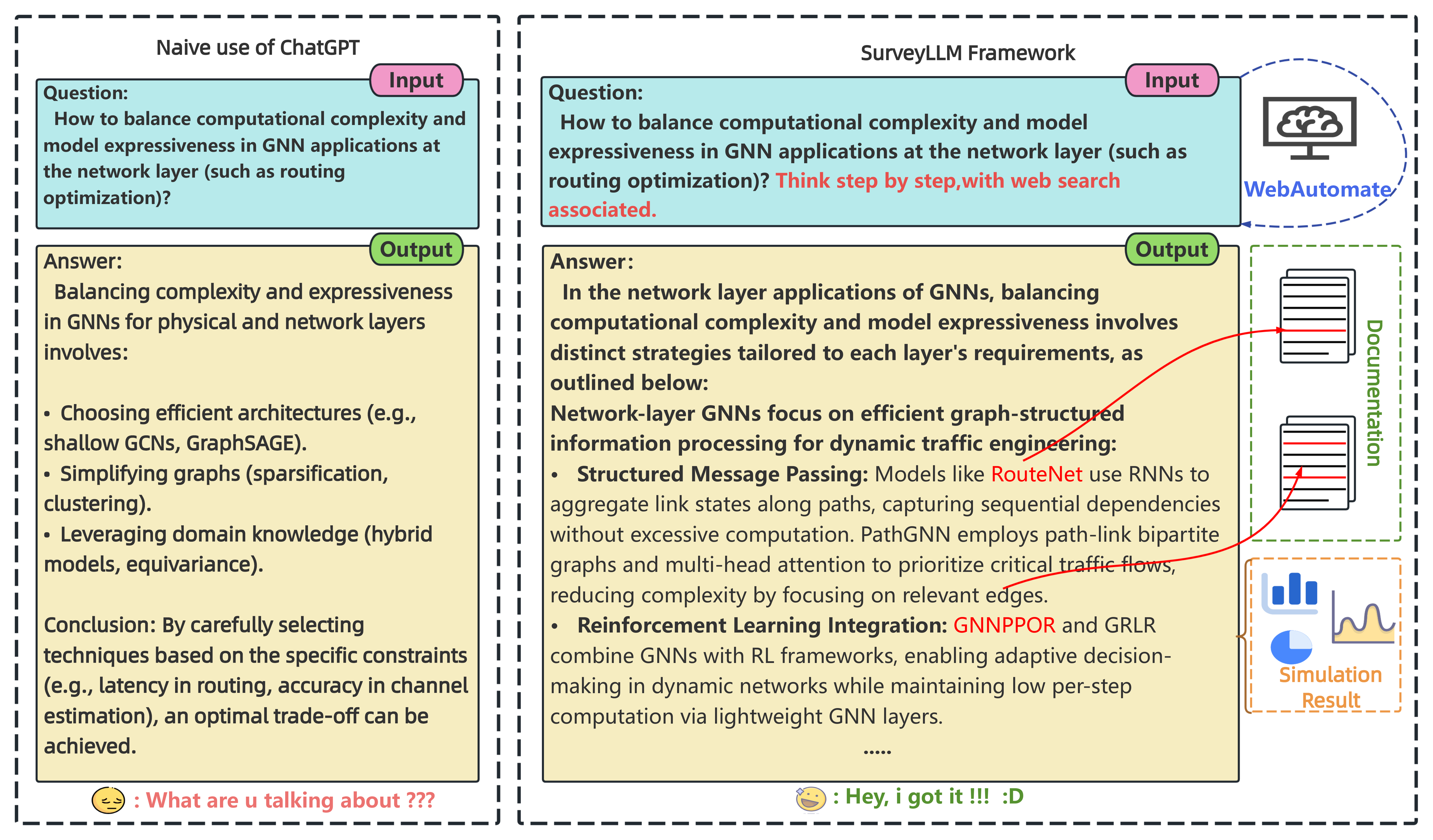}}
	\caption{The framework and case study of SurveyLLM.}
	\label{SurveyLLM} 
\end{figure}

\subsection{Case Study}

We conduct a case study to compare  vanilla  ChatGPT and SurveyLLM  in addressing questions regarding GNN-enabled wireless communications and networking, as illustrated in Figure \ref{SurveyLLM}(b). By leveraging the local domain knowledge base and RAG framework, SurveyLLM delivers domain-adaptive and accurate responses when addressing domain-specific queries (e.g., \texttt{"GNN applications at the network
layer"}). Specifically, vanilla ChatGPT offers overly general recommendations (e.g., \texttt{"select efficient architectures"}), whereas SurveyLLM delivers domain-specific, literature-grounded insights. These include citing specific models (e.g., RouteNet-F and HetGNNs), quantifying performance gains (e.g., a $43.78\%$ improvement), and integrating graph concepts (e.g., bipartite graphs and RNN-based aggregation). 

Once the readers input their queries, three data streams are involved in prompt generation as shown in Figure \ref{SurveyLLM}(a). We highlight two of these streams marked in blue and purple along with the RAG processes, as follows.
\begin{enumerate}
    \item \textbf{Label-based Classification (marked in blue):} The input queries are assigned  labels (e.g. network layer, routing optimization) by the classification model. The  document fragments with similar labels are retrieved, and then processed and selected to form the final response.

    \item \textbf{Semantic Matching (marked in purple):} The original queries also undergo the embedding-based vectorization. The vectors generated in this step capture the contextual relationships between words, enabling SurveyLLM to grasp  the semantic meaning of the queries. By matching query vectors against chunks in the vector database, the chunks (containing the keywords from the query) with the highest matching scores are retrieved.

    \item \textbf{RAG Execution:} The original queries, the most relevant  document fragments,  and the most closely matched chunks are assembled into a structured context window, and fed into the prompt‑engineering module. The generated  prompts enable LLMs to produce domain-specific responses that are both highly relevant and richly contextualized.

\end{enumerate}

In summary, the SurveyLLM framework can be regarded as an LLM-driven wireless application, serving as a complement to agentic GNNs. Particularly, agentic GNNs autonomously organize GNNs to enable multi-step and goal-directed decisions, while SurveyLLM assist operators in learning about the functions and interactions of  GNNs. Notably, SurveyLLM is generalizable to other survey articles, offering a novel paradigm for quick and comprehensive access to domain knowledge. With the explosive growth of research on wireless communications and networking, SurveyLLM is well-positioned to mitigate the challenges of knowledge fragmentation, enabling more efficient synthesis, comparison, and dissemination of cutting-edge insights in the field.


\section{Open Problems and Future Directions}

GNNs for wireless AI remain at their early stage. We outline the following open problems and future directions:

\subsection{Graph Representation for Complicated Wireless Networks}

Diverse graph representation methods have been proposed for wireless networks, where even identical wireless topologies can yield distinct graph structures. For instance, for an  MU-MISO network,  some works, like \cite{Zhang_twc_25}, represent antennas/BSs and users as nodes to form a homogeneous graph, while other works, like \cite{li_twc_25}, represent antenna-user links as nodes to form a  heterogeneous graph. Besides, initial node and edge features can be assigned using network parameters or set manually. Two primary methods address heterogeneity in wireless networks: 1) leveraging heterogeneous graph representations, and 2) unifying different entity types into a homogeneous framework. While complex graph representations enable fine-grained network feature extraction, they often compromise training efficiency.  

Graph representations define how interactions among network entities are captured via GNN layers and which features are updated to yield target solutions. The diversity of wireless scenarios and applications necessitates topology-aware and objective-oriented graph representations to facilitate effective network modeling and optimization. Designing and evaluating graph representation methods play a fundamental role in GNN-empowered wireless communications and networking.

\subsection{Powerful Architecture of GNN for Wireless Networks}

A growing number of GNN models have been reported for wireless communications and networks, and they enhance model expressiveness through two key aspects. First, improving the alignment between GNN architectures and wireless tasks is essential, necessitating a balance between inference efficiency and effectiveness. Additionally, the over-smoothing issue, which hinders the stacking of deeper GNN layers, remains a critical challenge. Second, integrating prior knowledge into GNN frameworks should span the entire architecture design, including input augmentation, layer processing, and output reformulation. Moreover,  the  flat networking architecture demands  efficient distributed implementation strategies for  GNNs, such as  OTA computation \cite{Yang_wcm_23}.

The design of neural architectures remains a central research focus across all DL applications. Currently, most existing GNN models for wireless applications are derived from frameworks (cf. Section II-B) originally  designed for other domains. Few studies have proposed GNN architectures specifically tailored to  wireless communications and networking. The in-depth alignment between GNN architectures and wireless tasks, along  with the integration of GNN models with domain-specific prior knowledge, constitutes a pivotal research direction.  

\subsection{Efficient Schemes for Complicated Constraints}

As discussed in Section II-D, an optimization problem solving approach is essential to obtain feasible solutions.  There are three methods to enforce the outputs of DL models to satisfy constraints: 1) scaling operation or activation functions; 2)  loss functions with penalty or Lagrangian regularization terms; 3) customized modules. As reported, simple-form constraints, such as  power budgets, $0$-$1$ integer LP, and semi-definite outputs, can be guaranteed via scaling or activation functions. Besides, most existing DL models incorporate regularization terms into loss functions to refine their outputs to satisfy complex constraints, such as QoS requirements. However, this approach often struggles to generate feasible solutions when confronted with tight constraints, thereby undermining the practical usability of the models. Moreover, only a few works have developed customized modules to project outputs onto feasible solution spaces. For example, a projection optimization problem is formulated and solved in \cite{Zhao_twc_25} as a post-processing module for DL outputs. 


\subsection{Model Robustness Maintenance}

The robustness challenges in GNN-based schemes entail two key dimensions. First, leveraging GNN-based schemes for robust signal processing. As elaborated on Section III-A, some studies propose GNN-enabled robust transmission designs under imperfect CSI scenarios. Second, unlike theory-driven methods, GNN-based schemes are data-driven, rendering them susceptible to adversarial manipulations and perturbations. For example, in \cite{Ghasemi_tvt_24}, the  authors first propose adversarial attacks to perturb a trained GNN model (originally proposed in \cite{Shen_jsac_21} for radio resource management in D2D communications) during the test phase, aiming to degrade the communication performance. Then, they propose an attack detection mechanism by analyzing the changes in the distribution of channel eigenvalues caused by attacks. 

Interpretability remains a critical challenge for DL-powered applications, particularly in communication systems where information transmission is exchanged across uncontrolled and time-varying channel environments, and among heterogeneous devices. Thus, enhancing model  robustness against noisy data, adversarial perturbations, and privacy risks \cite{10210275} becomes the central consideration for practical deployment.

\subsection{GNN-Enabled Wireless Application Layer}

Most existing studies employ GNNs at or below the network layer. However, GNNs have achieved success in certain domains as model augmenters, such as Graph Transformer \cite{Graph_Transformer} and Graph RAG \cite{edge_graphrag_0801}. Within the wireless application layer, GNNs can extract topology-aware features or capture interactions among mobile users to support downstream applications, including the generation of virtual system configurations (e.g., DT) and the analysis of user behaviors. Notably, in the field of AIGC, enhancing the wireless application layer could foster novel commercial models, thereby generating profits for telecom operators. Furthermore, with the proliferation of GNN-enabled applications, agentic GNNs can be further enhanced to handle more complex tasks, enabling the emergence of new service paradigms \cite{xiao2025agenticainetworking6g}.


\subsection{Dataset, Baselines, and Evaluation}

Unlike CVXopt-based algorithms and system performance analysis, which pursue  theoretical insights, DL models are engineered for practical implementation. However, most existing DL models focus on theoretical problems (rather than practical tasks) in wireless communications and networking. The lack of standardized datasets, well-defined task specifications, and uniform evaluation metrics impedes the reproducibility and inter-study comparison of experimental procedures in DL-enabled wireless signal processing. Therefore, integrating GNNs into wireless networks necessitates an in-depth exploration of the synergy between DL methodologies and wireless task characteristics. For instance, the GNN-enabled LDPC channel decoding scheme proposed in \cite{Cammerer_gc_22} is trained on standard-compliant datasets and evaluated against practical performance metrics.  Such alignment will expedite the deployment of GNNs in real-world systems.


\section{Conclusion}
This article has proposed agentic GNNs to enable scenario- and task-aware implementation of GNN models in wireless communications and networking towards EGI. We have presented a comprehensive survey of recent applications of GNNs designed for wireless systems. Specifically, we have focused on the alignment between graph representations and network topologies, as well as the alignment between neural architectures  and  wireless tasks. We have provided an overview of graph representation learning to clarify the tasks  that GNNs excel at addressing, introduced popular GNN models with their detailed architectures, and offered the framework of agentic GNNs. Then, we have summarized the applications of GNNs for traditional communication  and  networking intelligence, multifunctional services and diverse transmission, and flexible antennas and CFmMIMO, aiming to  analyze the key points to enhance the expressive capability of GNNs for  wireless communications and networking.  Finally, we have proposed an LLM framework as an intelligent  agent to answer technical questions, leveraging a knowledge base constructed from this article and its references.

\bibliographystyle{IEEEtran}
\bibliography{IEEEabrv,ref}

\end{document}